\documentclass{CEAS_EuroGNC_2024}

\usepackage[utf8]{inputenc}

\usepackage{graphicx}
\usepackage{amsmath}
\usepackage[version=4]{mhchem}
\usepackage{siunitx}
\usepackage{array}
\usepackage{longtable}
\usepackage{float}
\title{Singularity and Error Analysis of a Simple Quaternion Estimator}

\ifpdf
\hypersetup{pdfinfo={
Title={Guidelines and Example for the Preparation of the CEAS EuroGNC Conference Papers},
Author={Nicolas Fezans},
Subject={Template and Authors Instructions for CEAS EuroGNC Conferences},
Keywords={CEAS EuroGNC; Paper Preparation Guidelines; Authors Instructions; Template}
}}
\fi

\begin{document}
\maketitle

\renewcommand{\include}{\input}
\newcommand{\beq}{\begin{equation}}
\newcommand{\eeq}{\end{equation}}
\newcommand{\bdm}{\begin{displaymath}}
\newcommand{\edm}{\end{displaymath}}

\newcommand{\ever}{\ensuremath{\widetilde{\bf \theta}}}
\newcommand{\mever}{\ensuremath{\overline{\ever}}}
\newcommand{\sever}{\ensuremath{\sigma\ever}}

\newcommand{\mlmx}{\ensuremath{\overline{\lambda_{max}}}}
\newcommand{\slmx}{\ensuremath{\sigma_{\lmx}}}

\newcommand{\ger}{\ensuremath{\tilde{\bf{y}}}}

\newcommand{\Zer}{\ensuremath{\widetilde{\bf{Z}}}}

\newcommand{\sa}{\ensuremath{\sin{\alpha}}}
\newcommand{\ca}{\ensuremath{\cos{\alpha}}}
\newcommand{\sib}{\ensuremath{\sin{\beta}}}
\newcommand{\cb}{\ensuremath{\cos{\beta}}}

\newcommand{\Bfr}{\ensuremath{\mathcal{B}}}
\newcommand{\Btfr}{\ensuremath{\Bfr{_{t}}}}
\newcommand{\Befr}{\ensuremath{  \widehat{\Bfr} }}
\newcommand{\Bfrk}{\ensuremath{ \Bfr_{_k} }}
\newcommand{\Bfri}{\ensuremath{ \Bfr_{_i} }}
\newcommand{\Cfr}{\ensuremath{\mathcal{C}}}
\newcommand{\Rfr}{\ensuremath{\mathcal{R}}}
\newcommand{\Ifr}{\ensuremath{\mathcal{I}}}

\newcommand{\eqdef}{\ensuremath{{\;\stackrel{\triangle}{=}\;}}}

\newtheorem{lem}{Lemma}
\newtheorem{defi}{Definition}

\newcommand{\adjoint}{\ensuremath{ \,\mbox{adj} }}
\newcommand{\const}{\ensuremath{ \,\mbox{const}   }}
\newcommand{\cov}{\ensuremath{ \,\mbox{cov}   }}
\newcommand{\dghr}{\ensuremath{ \,\mbox{$\frac{deg}{hr}$} }}
\newcommand{\dt}{\ensuremath{ \,\Delta t   }}
\newcommand{\dst}{\ensuremath{ \, dt   }}
\newcommand{\eqr}{Eq.~\eqref }
\newcommand{\expm}{\ensuremath{ \mbox{\emph{\Large e}}} }
\newcommand{\half}{\ensuremath{ \,\frac{1}{2}\,   }}
\newcommand{\smallhalf}{\ensuremath{ \texttt{\small\half} }}
\newcommand{\quarter}{\ensuremath{ \,\frac{1}{4}\,   }}
\newcommand{\smallquarter}{\ensuremath{ \texttt{\small\quarter} }}
\newcommand{\halfdt}{\ensuremath{ \,\frac{\dt}{2}\,   }}
\newcommand{\kernel}{\ensuremath{ \,\mbox{Ker} }}
\newcommand{\mdghr}{\ensuremath{ \,\mbox{mdeg/hr} }}
\newcommand{\Order}{\ensuremath{ \,\mathcal{O} }}
\newcommand{\rank}{\ensuremath{ \,\mbox{rank} }}
\newcommand{\real}{\ensuremath{ \mathbb{R} }}
\newcommand{\realthree}{\ensuremath{ \real^{3} }}
\newcommand{\realfour}{\ensuremath{ \real^{4} }}
\newcommand{\realsixteen}{\ensuremath{ \real^{16} }}
\newcommand{\realn}{\ensuremath{ \real^{n} }}
\newcommand{\realm}{\ensuremath{ \real^{m} }}
\newcommand{\realfourbyfour}{\ensuremath{ \real^{4\times 4} }}
\newcommand{\realnbym}{\ensuremath{ \real^{n\times m} }}
\newcommand{\remo}{ \emph{Remark 1}: }
\newcommand{\remtw}{ \emph{Remark 2}: }
\newcommand{\remth}{ \emph{Remark 3}: }
\newcommand{\remf}{ \emph{Remark 4}: }
\newcommand{\Span}{ \mbox{Span} }
\newcommand{\sigbyfor}{\ensuremath{ \frac{\sig^2}{4} }}
\newcommand{\smallsigbyfor}{\ensuremath{ \texttt{\small\sigbyfor} }}
\newcommand{\sigepsbyfor}{\ensuremath{ \frac{\sigeps^2}{4} }}
\newcommand{\smallsigepsbyfor}{\ensuremath{ \texttt{\small\sigepsbyfor} }}
\newcommand{\spectrum}{\ensuremath{ \,\mbox{Sp} }}
\newcommand{\thrsigbyfor}{\ensuremath{ \frac{3\sig^2}{4} }}
\newcommand{\smallthrsigbyfor}{\ensuremath{ \texttt{\small\thrsigbyfor} }}
\newcommand{\thrsigepsbyfor}{\ensuremath{ \frac{3\sigeps^2}{4} }}
\newcommand{\smallthrsigepsbyfor}{\ensuremath{ \texttt{\small\thrsigepsbyfor} }}
\newcommand{\trace}{\ensuremath{ \,\mbox{tr} }}
\newcommand{\vctr}{\ensuremath{ \,\mbox{vec}   }}

\newcommand{\inm}{\ensuremath{{Q}}}
\newcommand{\onm}{\ensuremath{{R}}}
\newcommand{\pdm}{\ensuremath{P^d }}
\newcommand{\pttm}{\ensuremath{ P^{\xi} }}
\newcommand{\rttm}{\ensuremath{ \mathcal{Y}^{\xi} }}
\newcommand{\onttm}{\ensuremath{ \onm^{\xi} }}
\newcommand{\kttm}{\ensuremath{ K^{\xi} }}


\newcommand{\av}{\ensuremath{ {\bf a}  }}
\newcommand{\ahat}{\ensuremath{ \widehat{\av} }}
\newcommand{\ahatv}{\ensuremath{ \widehat{\av} }}

\newcommand{\bv}{\ensuremath{ \,{\bf b}  }}
\newcommand{\bonev}{\ensuremath{ \bv_{_{1}} }}
\newcommand{\btwov}{\ensuremath{ \bv_{_{2}} }}
\newcommand{\biv}{\ensuremath{ \bv_{_{i}} }}
\newcommand{\bi}{\ensuremath{ \,{\bf b}_{_i}  }}
\newcommand{\bthrv}{\ensuremath{ \bv_{_{3}} }}
\newcommand{\bion}{\ensuremath{ \bi^{^1} }}
\newcommand{\bitw}{\ensuremath{ \bi^{^2} }}
\newcommand{\bko}{\ensuremath{ \,{\bf b}_{_{k+1}}  }}
\newcommand{\bkoi}{\ensuremath{ \bko^i }}
\newcommand{\bk}{\ensuremath{ \,{\bf b}_{_k}  }}
\newcommand{\bki}{\ensuremath{ \,{\bf b}_{_{k+i}}  }}
\newcommand{\bz}{\ensuremath{ {\,{\bf b}^o}  }}
\newcommand{\bkt}{\ensuremath{  \bz_{_k} }}
\newcommand{\bkot}{\ensuremath{  \bz_{_{k+1}} }}
\newcommand{\bt}{\ensuremath{ \bz }}
\newcommand{\bo}{\ensuremath{ {\bf b}_{_1}  }}
\newcommand{\btwo}{\ensuremath{ {\bf b}_{_2}  }}
\newcommand{\bn}{\ensuremath{ {\bf b}_{_n}  }}
\newcommand{\berv}{\ensuremath{\mathbf{\delta b}}}
\newcommand{\bers}{\ensuremath{{\sigma_b}}}
\newcommand{\bq}{\ensuremath{ \bv_{_q} }}
\newcommand{\be}{\ensuremath{ \widehat{\bv} }}
\newcommand{\bei}{\ensuremath{ \be_{_{i}} }}

\newcommand{\betav}{\ensuremath{ \,{\boldsymbol\beta}  }}
\newcommand{\betatv}{\ensuremath{ \betav_{_{\!t}} }}
\newcommand{\betativ}{\ensuremath{ \betav_{_{t_i}} }}
\newcommand{\betatov}{\ensuremath{ \betav_{_{\!\tau}} }}

\newcommand{\bital}{\ensuremath{ \emph{\bv} }}

\newcommand{\dbv}{\ensuremath{{\,\boldsymbol{\delta\!b}}}}
\newcommand{\dbi}{\ensuremath{ \dbv_{_{i}}  }}
\newcommand{\dbj}{\ensuremath{ \dbv_{_{j}}  }}
\newcommand{\dbk}{\ensuremath{ \dbv_{_{\!k}}  }}
\newcommand{\dbki}{\ensuremath{ \dbv_{_{k+i}}  }}
\newcommand{\dbko}{\ensuremath{ \dbv_{_{k+1}}  }}
\newcommand{\dbkoi}{\ensuremath{ {\dbko^i} }}
\newcommand{\dbkoj}{\ensuremath{ {\dbko^j} }}

\newcommand{\dv}{\ensuremath{{\bf d}}}
\newcommand{\dkv}{\ensuremath{ \dv_{_{k}} }}
\newcommand{\dk}{\ensuremath{ \dv_{_{k}} }}
\newcommand{\donev}{\ensuremath{ \dv_{_{1}} }}
\newcommand{\dtwov}{\ensuremath{ \dv_{_{2}} }}
\newcommand{\djv}{\ensuremath{ \dv_{_{j}} }}
\newcommand{\dko}{\ensuremath{ \dv_{_{k+1}} }}
\newcommand{\dumykk}{\ensuremath{  {\widehat{\left( \bullet \right)}}_{_{k/k}}    }}
\newcommand{\dumykok}{\ensuremath{  {\widehat{\left( \bullet \right)}}_{_{k+1/k}}    }}
\newcommand{\dumyerkk}{\ensuremath{  {\widetilde{\left( \bullet \right)}}_{_{k/k}}    }}
\newcommand{\dumyerkok}{\ensuremath{  {\widetilde{\left( \bullet \right)}}_{_{k+1/k}}    }}

\newcommand{\Dv}{\ensuremath{{\bf \Delta}}}
\newcommand{\Dbv}{\ensuremath{{\Dv\!\bv }}}
\newcommand{\Dbiv}{\ensuremath{{\Dv\!\bv_i }}}
\newcommand{\Driv}{\ensuremath{{\Dv\!\rv_i }}}
\newcommand{\PDbjv}{\ensuremath{{P_{\Dv\bv_j }}}}
\newcommand{\PDrjv}{\ensuremath{{P_{\Dv\rv_j }}}}
\newcommand{\Dbonev}{\ensuremath{{\Dv\!\bonev }}}
\newcommand{\Dbtwov}{\ensuremath{{\Dv\!\btwov }}}
\newcommand{\Drv}{\ensuremath{{\Dv\!\rv }}}
\newcommand{\Dronev}{\ensuremath{{\Dv\!\ronev }}}
\newcommand{\Drtwov}{\ensuremath{{\Dv\!\rtwov }}}
\newcommand{\Dqbar}{\ensuremath{{\Dv\!\qbar }}}
\newcommand{\Ddonev}{\ensuremath{{\Dv\!\donev }}}
\newcommand{\Ddtwov}{\ensuremath{{\Dv\!\dtwov }}}
\newcommand{\Dsonev}{\ensuremath{{\Dv\!\sonev }}}
\newcommand{\Dstwov}{\ensuremath{{\Dv\!\stwov }}}
\newcommand{\Ddv}{\ensuremath{{\Dv\!\dv }}}
\newcommand{\Dsv}{\ensuremath{{\Dv\!\sv }}}
\newcommand{\Ddiv}{\ensuremath{{\Dv\!\dv_i }}}
\newcommand{\Dsiv}{\ensuremath{{\Dv\!\sv_i }}}
\newcommand{\Dqsca}{\ensuremath{{\Dv\!\qsca }}}

\newcommand{\dev}{\ensuremath{{\,\boldsymbol{\delta e}}}}
\newcommand{\dqv}{\ensuremath{\boldsymbol{\delta q}  }}
\newcommand{\dqvs}{\ensuremath{\delta q}}
\newcommand{\dqkk}{\ensuremath{ \dqv_{_{k/k}} }}
\newcommand{\dqkok}{\ensuremath{ \dqv_{_{k+1/k}} }}

\newcommand{\dqtv}{\ensuremath{{\bf \dv\!\qtv}}}
\newcommand{\dqetv}{\ensuremath{{\bf \dv\!\qetv}}}
\newcommand{\dbetav}{\ensuremath{{\bf \dv\!\!\betav}}}
\newcommand{\dbetatv}{\ensuremath{ \dbetav_t }}
\newcommand{\dbetatov}{\ensuremath{{\bf \dv\!\!\betatov}}}
\newcommand{\dztv}{\ensuremath{{\bf \dv\ztv}}}
\newcommand{\dnutv}{\ensuremath{{\bf \dv\!\nutv}}}
\newcommand{\dxtv}{\ensuremath{{\bf \dv\!\xtv}}}
\newcommand{\dqtldtv}{\ensuremath{{\bf \dv\!\qtldtv}}}
\newcommand{\dntv}{\ensuremath{{\bf \dv\!\ntv}}}
\newcommand{\dytv}{\ensuremath{{\bf \dv\!\ytv}}}

\newcommand{\deltav}{\ensuremath{\boldsymbol{\delta} }}
\newcommand{\delqbar}{\ensuremath{{\deltav\!\qbar }}}
\newcommand{\delqev}{\ensuremath{{\deltav\!\qev }}}

\newcommand{\deltetav}{\ensuremath{{\deltav\!\tetav }}}

\newcommand{\Dqv}{\ensuremath{\boldsymbol{\Delta\!q} }}
\newcommand{\Dqkk}{\ensuremath{\Dqv_{_{k/k}} }}
\newcommand{\Dqkok}{\ensuremath{ \Dqv_{_{k+1/k}} }}
\newcommand{\Dqz}{\ensuremath{ \Dqv^0 }}
\newcommand{\Dqzkok}{\ensuremath{ \Dqz_{_{k+1/k}} }}
\newcommand{\Dqi}{\ensuremath{ \Dqv^i }}
\newcommand{\Dqikok}{\ensuremath{ \Dqi_{_{k+1/k}} }}
\newcommand{\Dqev}{\ensuremath{\boldsymbol{\Delta \! \qev} }}

\newcommand{\ev}{\ensuremath{ \,{\bf e}  }}
\newcommand{\etz}{\ensuremath{ \ev^0 }}
\newcommand{\ek}{\ensuremath{ \ev_{_k} }}
\newcommand{\eko}{\ensuremath{ \ev_{_{k+1}} }}
\newcommand{\ee}{\ensuremath{ \widehat{\ev}  }}
\newcommand{\eek}{\ensuremath{ \ee_{_{k}} }}
\newcommand{\eekk}{\ensuremath{ \ee_{_{k/k}} }}
\newcommand{\eeko}{\ensuremath{ \ee_{_{k+1}} }}
\newcommand{\ekok}{\ensuremath{ \ee_{_{k+1/k}}  }}
\newcommand{\eekok}{\ensuremath{ \ee_{_{k+1/k}}  }}
\newcommand{\ekoko}{\ensuremath{ \ee_{_{k+1/k+1}}  }}
\newcommand{\eekoko}{\ensuremath{ \ee_{_{k+1/k+1}}  }}
\newcommand{\ei}{\ensuremath{ \,{\bf e}_{_i}  }}
\newcommand{\eekolk}{\ensuremath{ \ee_{_{k+l-1/k}}  }}
\newcommand{\eeklk}{\ensuremath{ \ee_{_{k+l/k}}  }}

\newcommand{\eulv}{\ensuremath{ \,\boldsymbol{\theta} }}

\newcommand{\epsv}{\ensuremath{ \,\boldsymbol{\epsilon} }}
\newcommand{\epsk}{\ensuremath{ \epsv_{_k} }}
\newcommand{\epski}{\ensuremath{ \epsv_{_{k+i}} }}
\newcommand{\epsi}{\ensuremath{  \epsv_{_i} }}
\newcommand{\epsko}{\ensuremath{ \epsv_{_{k+1}} }}
\newcommand{\epsok}{\ensuremath{ \epsv_{_{k-1}} }}
\newcommand{\epsktt}{\ensuremath{  \epsv_{k,\,true} }}
\newcommand{\epsbar}{\ensuremath{ \overline{\epsv}  }}
\newcommand{\epsbark}{\ensuremath{ \epsbar_{_k}  }}
\newcommand{\epskbar}{\ensuremath{ \epsbar_{_k}  }}
\newcommand{\epse}{\ensuremath{ \widehat{\epsv} }}
\newcommand{\epseokk}{\ensuremath{ \epse_{_{k-1/k}} }}
\newcommand{\epsekk}{\ensuremath{ \epse_{_{k/k}} }}
\newcommand{\epstv}{\ensuremath{ \epsv{_{_t}} }}
\newcommand{\epstov}{\ensuremath{ \epsv{_{_\tau}} }}

\newcommand{\epsxtk}{\ensuremath{ \epsk^{16} }}

\newcommand{\fv}{\ensuremath{ \,{\bf f}  }}

\newcommand{\fik}{\ensuremath{{ {\bf \varphi}_{_k}}}}

\newcommand{\inv}{\ensuremath{{{\bf w}}}}

\newcommand{\minv}{\ensuremath{\overline{\inv}}}
\newcommand{\monv}{\ensuremath{\overline{\onv}}}
\newcommand{\mberv}{\ensuremath{\mathbf{\mu_b}}}
\newcommand{\mr}{\ensuremath{{\bf{m}}_r}}
\newcommand{\mepsk}{\ensuremath{ \mv_{_{\eps_{k}}}    }}

\newcommand{\nv}{\ensuremath{ {\bf n}  }}
\newcommand{\no}{\ensuremath{  \nv_{_1} }}
\newcommand{\nt}{\ensuremath{   \nv_{_2}  }}
\newcommand{\nepsk}{\ensuremath{ \nv_{_{\eps_{k}}}    }}
\newcommand{\nwk}{\ensuremath{ \nv_{_{w_{k}}}    }}
\newcommand{\ntv}{\ensuremath{ \nv_{_{t}} }}
\newcommand{\ntov}{\ensuremath{ \nv_{_{\tau}} }}

\newcommand{\nuv}{\ensuremath{ {\boldsymbol \nu}  }}
\newcommand{\nutv}{\ensuremath{ \nuv_{_{t}} }}

\newcommand{\Ov}{\ensuremath{ {\bf{0}}  }}
\newcommand{\Onev}{\ensuremath{ {\bf{1}}  }}

\newcommand{\omg}{\ensuremath{ \,\boldsymbol{\omega} }}
\newcommand{\omgv}{\ensuremath{ \,\boldsymbol{\omega} }}
\newcommand{\omk}{\ensuremath{  \omgv_{_k}  }}
\newcommand{\omkt}{\ensuremath{ {\omgv_{_k}^{o}}  }}
\newcommand{\omok}{\ensuremath{ \omgv^0_{_k} }}
\newcommand{\ome}{\ensuremath{  \widehat{\omgv} }}
\newcommand{\omekk}{\ensuremath{  \ome_{_{k/k}}  }}
\newcommand{\omgtv}{\ensuremath{  \omgv_{_t}  }}
\newcommand{\omgtzv}{\ensuremath{  \omgtv^{o}  }}
\newcommand{\omgtzvcross}{\ensuremath{ \,\left[ \omgtzv \times \right] }}

\newcommand{\onv}{\ensuremath{{{\bf v}}}}

\newcommand{\qv}{\ensuremath{{\bf q}}}
\newcommand{\qt}{\ensuremath{ \qv_{_t}}}
\newcommand{\qtv}{\ensuremath{ \qv_{_t} }}
\newcommand{\qonev}{\ensuremath{{\qv_{_{1}}}}}
\newcommand{\qtwov}{\ensuremath{{\qv_{_{2}}}}}
\newcommand{\qthrv}{\ensuremath{{\qv_{_{3}}}}}
\newcommand{\qforv}{\ensuremath{{\qv_{_{4}}}}}
\newcommand{\qonethrv}{\ensuremath{{\qv_{_{13}}}}}
\newcommand{\qoneforv}{\ensuremath{{\qv_{_{14}}}}}
\newcommand{\qtwothrv}{\ensuremath{{\qv_{_{23}}}}}
\newcommand{\qtwoforv}{\ensuremath{{\qv_{_{24}}}}}
\newcommand{\qiv}{\ensuremath{{\qv_{_{i}}}}}

\newcommand{\qijv}{\ensuremath{{\qv_{_{ij}}}}}
\newcommand{\qtz}{\ensuremath{ {\bf q}^0  }}
\newcommand{\qto}{\ensuremath{ {\bf q}^1  }}
\newcommand{\qtw}{\ensuremath{ {\bf q}^2  }}
\newcommand{\qth}{\ensuremath{ {\bf q}^3  }}
\newcommand{\qk}{\ensuremath{{  {\bf q}_{_k}}}}
\newcommand{\qak}{\ensuremath{{  {\bf q}_{_k}}}}
\newcommand{\qko}{\ensuremath{ \qv_{_{k+1}} }}
\newcommand{\qok}{\ensuremath{ \qv_{_{k-1}} }}
\newcommand{\qz}{\ensuremath{{  {\bf q}_{_0}}}}
\newcommand{\qon}{\ensuremath{ \qv_{_{1}} }}
\newcommand{\qtwo}{\ensuremath{ \qv_{_{2}} }}
\newcommand{\qN}{\ensuremath{ \qv_{_N} }}
\newcommand{\qoN}{\ensuremath{ \qv_{_{N-1}} }}
\newcommand{\qNo}{\ensuremath{ \qv_{_{N+1}} }}

\newcommand{\qiko}{\ensuremath{ {\qko^i} }}
\newcommand{\qzko}{\ensuremath{ {\qko^0} }}
\newcommand{\qoko}{\ensuremath{ {\qko^1} }}
\newcommand{\qtwko}{\ensuremath{ {\qko^2} }}
\newcommand{\qthko}{\ensuremath{ {\qko^3} }}

\newcommand{\qkt}{\ensuremath{ \qk^{true} }}
\newcommand{\qkot}{\ensuremath{ \qko^{true} }}
\newcommand{\qzt}{\ensuremath{ \qv_0^{true} }}

\newcommand{\qev}{\ensuremath{ \widehat{\bf q} }}
\newcommand{\qe}{\ensuremath{ \widehat{\bf q} }}
\newcommand{\qek}{\ensuremath{ \qe_{_k} }}
\newcommand{\qeko}{\ensuremath{ \qe_{_{k+1}} }}
\newcommand{\qekk}{\ensuremath{   \qe_{_{k/k}} }}
\newcommand{\qeNN}{\ensuremath{   \qe_{_{N/N}} }}
\newcommand{\qeNoN}{\ensuremath{   \qe_{_{N+1/N}} }}
\newcommand{\qeKok}{\ensuremath{  \qe_{_{k/k-1}}  }}
\newcommand{\qeNon}{\ensuremath{  \qe_{_{N/N-1}}  }}
\newcommand{\qeoNoN}{\ensuremath{  \qe_{_{N-1/N-1}}  }}
\newcommand{\qeNoNo}{\ensuremath{   \qe_{_{N+1/N+1}} }}
\newcommand{\qekoko}{\ensuremath{ \qe_{_{k+1/k+1}} }}
\newcommand{\qeokok}{\ensuremath{ \qe_{_{k-1/k-1}} }}
\newcommand{\qekokoast}{\ensuremath{ \qekoko^{\ast} }}
\newcommand{\qekkast}{\ensuremath{ \qekk^{\ast} }}
\newcommand{\qekok}{\ensuremath{   \qe_{_{k+1/k}} }}
\newcommand{\qeklk}{\ensuremath{   \qe_{_{k+l/k}} }}
\newcommand{\qekolk}{\ensuremath{   \qe_{_{k+l-1/k}} }}
\newcommand{\qezz}{\ensuremath{    \qe_{_{0/0}} }}
\newcommand{\qeoz}{\ensuremath{    \qe_{_{1/0}} }}
\newcommand{\qeoo}{\ensuremath{     \qe_{_{1/1}} }}
\newcommand{\qzstar}{\ensuremath{{ {\star{\bf q}}_{_0} }}}
\newcommand{\qostar}{\ensuremath{{ {\star{\bf q}}_{_1} }}}
\newcommand{\qater}{\ensuremath{\tilde{\bf q}}}
\newcommand{\qaterkoko}{\ensuremath{ \qater_{_{k+1/k+1}}    }}
\newcommand{\qeki}{\ensuremath{{ \,\widehat{\bf q}_{_{k/i} }}}}

\newcommand{\qetv}{\ensuremath{  \qe_{_{t}} }}

\newcommand{\qezkok}{\ensuremath{ {\qekok^0} }}
\newcommand{\qeonekok}{\ensuremath{ {\qekok^1} }}
\newcommand{\qetwkok}{\ensuremath{ {\qekok^2} }}
\newcommand{\qethkok}{\ensuremath{ {\qekok^3} }}
\newcommand{\qeikok}{\ensuremath{ {\qekok^i} }}

\newcommand{\qb}{\ensuremath{ \overline{\bf q}  }}
\newcommand{\qbar}{\ensuremath{  \,\bar{\bf q}  }}
\newcommand{\qbarz}{\ensuremath{  \qbar_{_0} }}
\newcommand{\qbaro}{\ensuremath{  \qbar_{_1} }}
\newcommand{\qbarN}{\ensuremath{  \qbar_{_N} }}
\newcommand{\qbaroN}{\ensuremath{  \qbar_{_{N-1}} }}
\newcommand{\qbark}{\ensuremath{  \qbar_{_k} }}
\newcommand{\qbarok}{\ensuremath{  \qbar_{_{k-1}} }}
\newcommand{\qbari}{\ensuremath{  \qbar_{_i} }}
\newcommand{\qbarKok}{\ensuremath{  \qbar_{_{k/k-1}}   }}
\newcommand{\qbarbar}{\ensuremath{      \bar{\qbar}  }}
\newcommand{\qbarbarKok}{\ensuremath{  \qbarbar_{_{k/k-1}} }}
\newcommand{\qsca}{\ensuremath{  \,\check{\mathbf{q}}}}

\newcommand{\qkbar}{\ensuremath{{  \bar{\bf q}_{_{k}} }}}
\newcommand{\qzbar}{\ensuremath{{  \bar{\bf q}_{_0} }}}
\newcommand{\qobar}{\ensuremath{{  \bar{\bf q}_{_1} }}}
\newcommand{\qkobar}{\ensuremath{{  \bar{\bf q}_{_{k+1}} }}}

\newcommand{\qkk}{\ensuremath{ {\bf q}_{_{k/k}}  }}
\newcommand{\qerkk}{\ensuremath{ {\boldsymbol{\delta}} \qkk }}
\newcommand{\qkok}{\ensuremath{ {\bf q}_{_{k+1/k}}  }}
\newcommand{\qerkok}{\ensuremath{ {\boldsymbol{\delta}} \qkok }}
\newcommand{\qkoko}{\ensuremath{ {\bf q}_{_{k+1/k+1}}  }}
\newcommand{\qerkoko}{\ensuremath{ {\boldsymbol{\delta}} \qkoko }}
\newcommand{\qerkokoast}{\ensuremath{ \qerkoko^{\ast} }}
\newcommand{\qierkok}{\ensuremath{ {\qerkok^i} }}
\newcommand{\qzerkok}{\ensuremath{ {\qerkok^0} }}

\newcommand{\qtldv}{\ensuremath{ \tilde{\bf q} }}
\newcommand{\qtldtv}{\ensuremath{ \qtldv_{_{t}} }}

\newcommand{\Qvo}{\ensuremath{  {\bf{e}}_{_1}  }}
\newcommand{\Qvtw}{\ensuremath{  {\bf{e}}_{_2}  }}
\newcommand{\Qvth}{\ensuremath{  {\bf{e}}_{_3}  }}
\newcommand{\Qv}{\ensuremath{  {\bf{e}}  }}
\newcommand{\Quatv}{\ensuremath{{\bf e}}}

\newcommand{\rv}{\ensuremath{ \,{\bf r}  }}
\newcommand{\ronev}{\ensuremath{ \rv_{_{1}} }}
\newcommand{\rtwov}{\ensuremath{ \rv_{_{2}} }}
\newcommand{\riv}{\ensuremath{ \rv_{_{i}} }}
\newcommand{\ri}{\ensuremath{ \,{\bf r}_{_i}  }}
\newcommand{\rthrv}{\ensuremath{ \rv_{_{3}} }}
\newcommand{\rj}{\ensuremath{ \,{\bf r}_{_j}  }}
\newcommand{\rko}{\ensuremath{ \rv_{_{k+1}} }}
\newcommand{\rkoi}{\ensuremath{ {\rv_{_{k+1}}^i} }}
\newcommand{\rz}{\ensuremath{ \,{\bf r}_{_0}  }}
\newcommand{\ro}{\ensuremath{ \,{\bf r}_{_1}  }}
\newcommand{\rt}{\ensuremath{ \,{\bf r}_{_2}  }}
\newcommand{\rtwo}{\ensuremath{ \,{\bf r}_{_2}  }}
\newcommand{\rn}{\ensuremath{ \,{\bf r}_{_n}  }}
\newcommand{\rttv}{\ensuremath{ \,\bf{\xi}_{r} }}
\newcommand{\res}{\ensuremath{ \,\widehat{\bf e}}}
\newcommand{\resid}{\ensuremath{ \,{\bf r}}}
\newcommand{\rkok}{\ensuremath{ \rv_{_{k+1/k}}  }}
\newcommand{\rk}{\ensuremath{ {\bf r}_{_k}  }}
\newcommand{\rvq}{\ensuremath{ \rv_{_q} }}
\newcommand{\rklk}{\ensuremath{ \rv_{_{k+l/k}}  }}
\newcommand{\rknk}{\ensuremath{ \rv_{_{k+n/k}}  }}
\newcommand{\rvo}{\ensuremath{ \rv_{_1} }}
\newcommand{\rvtw}{\ensuremath{ \rv_{_2} }}
\newcommand{\rvth}{\ensuremath{ \rv_{_3} }}
\newcommand{\rqv}{\ensuremath{ \rv^q }}
\newcommand{\rqko}{\ensuremath{ \rqv_{_{k+1}} }}

\newcommand{\sv}{\ensuremath{{\bf s}}}
\newcommand{\sko}{\ensuremath{ \sv_{_{k+1}} }}
\newcommand{\sk}{\ensuremath{ \sv_{_{k}} }}
\newcommand{\sonev}{\ensuremath{ \sv_{_{1}} }}
\newcommand{\stwov}{\ensuremath{ \sv_{_{2}} }}
\newcommand{\siv}{\ensuremath{ \sv_{_{i}} }}

\newcommand{\tetav}{\ensuremath{ \,\boldsymbol{\theta} }}
\newcommand{\tetavko}{\ensuremath{ \,\boldsymbol{\theta}_{_{k+1}} }}
\newcommand{\tetaev}{\ensuremath{ \,\widehat{\tetav} }}
\newcommand{\tetaeWLS}{\ensuremath{ \tetaev^{_{_{WLS}}} }}
\newcommand{\tetaeWLSk}{\ensuremath{ \tetaeWLS_{_k} }}
\newcommand{\tetaeWLSko}{\ensuremath{ \tetaeWLS_{_{k+1}} }}

\newcommand{\uv}{\ensuremath{\mathbf{u}}}
\newcommand{\ui}{\ensuremath{ \uv_{_i}   }}
\newcommand{\ujv}{\ensuremath{ \uv_{_j}   }}
\newcommand{\uonev}{\ensuremath{ \uv_{_1}   }}
\newcommand{\utwov}{\ensuremath{ \uv_{_2}   }}
\newcommand{\uthrv}{\ensuremath{ \uv_{_3}   }}
\newcommand{\uforv}{\ensuremath{ \uv_{_4}   }}
\newcommand{\uiv}{\ensuremath{ \uv_{_i}   }}

\newcommand{\uhat}{\ensuremath{ \widehat{\uv} }}

\newcommand{\Uqv}{\ensuremath{\mathbf{1_q}}}

\newcommand{\vvv}{\ensuremath{ \mathbf{v}  }}

\newcommand{\vk}{\ensuremath{  \vvv_{_{\!k}} }}
\newcommand{\vko}{\ensuremath{  \vvv_{_{k+1}} }}
\newcommand{\vkon}{\ensuremath{  \vko^n }}
\newcommand{\vkoi}{\ensuremath{  \vko^i }}
\newcommand{\vi}{\ensuremath{  \vvv_{_i} }}
\newcommand{\von}{\ensuremath{  \vvv_{_1} }}
\newcommand{\vtw}{\ensuremath{  \vvv_{_2} }}
\newcommand{\vth}{\ensuremath{  \vvv_{_3} }}

\newcommand{\vbarv}{\ensuremath{ {\overline{\vvv}} }}
\newcommand{\vbark}{\ensuremath{ \vbarv_{_{k}} }}

\newcommand{\wv}{\ensuremath{  \,{\bf w} }}
\newcommand{\wonev}{\ensuremath{  \wv_{_{1}} }}
\newcommand{\wtwov}{\ensuremath{  \wv_{_{2}} }}
\newcommand{\wk}{\ensuremath{  \,{\bf w}_{_k} }}
\newcommand{\wok}{\ensuremath{  \wv_{_{k-1}} }}
\newcommand{\wN}{\ensuremath{  {\bf w}_{_N} }}
\newcommand{\woN}{\ensuremath{  {\wv_{_{N-1}}}  }}
\newcommand{\wko}{\ensuremath{     \wv_{_{k+1}} }}
\newcommand{\wtwN}{\ensuremath{  \wv_{_{N-2}} }}
\newcommand{\wz}{\ensuremath{  {\bf w}_{_0}   }}
\newcommand{\wo}{\ensuremath{  {\bf w}_{_1}  }}
\newcommand{\wsxtk}{\ensuremath{  \wk^{16} }}
\newcommand{\wtenk}{\ensuremath{  \wk^{10} }}
\newcommand{\wnink}{\ensuremath{  \wk^{9} }}
\newcommand{\wi}{\ensuremath{  \wv_{_i} }}

\newcommand{\wbar}{\ensuremath{ \,\bar{\wv} }}
\newcommand{\wkbar}{\ensuremath{ \wbar_{_k} }}
\newcommand{\wbark}{\ensuremath{ \wbar_{_k} }}
\newcommand{\wbarok}{\ensuremath{ \wbar_{_{k-1}} }}
\newcommand{\wbarN}{\ensuremath{   \wbar_{_{N}} }}
\newcommand{\wbaroN}{\ensuremath{   \wbar_{_{N-1}} }}
\newcommand{\wkobar}{\ensuremath{     \wbar_{_{k+1}} }}
\newcommand{\wzbar}{\ensuremath{   \wbar_{_0} }}
\newcommand{\wbarz}{\ensuremath{   \wbar_{_0} }}
\newcommand{\wbartwk}{\ensuremath{  \wbar_{_{k-2}} }}
\newcommand{\wbartwN}{\ensuremath{  \wbar_{_{N-2}} }}
\newcommand{\wbarKok}{\ensuremath{  \wbar_{_{k/k-1}} }}
\newcommand{\wbarbar}{\ensuremath{  \bar{\wbar}  }}
\newcommand{\wbarbarokok}{\ensuremath{  \wbarbar_{_{k-1/k-1}}  }}

\newcommand{\wzstar}{\ensuremath{{  \star{\bf w}_{_0}}}}
\newcommand{\wostar}{\ensuremath{{  \star{\bf w}_{_1}}}}
\newcommand{\wstar}{\ensuremath{  \, {\wv}^{\star} }}
\newcommand{\wstarz}{\ensuremath{ \wstar_{_0}  }}
\newcommand{\wstaro}{\ensuremath{ \wstar_{_1}  }}
\newcommand{\wstarok}{\ensuremath{ \wstar_{_{k-1}}  }}
\newcommand{\wstarokqk}{\ensuremath{
 \wstarok \left( \qk \right) }}
\newcommand{\wstartwk}{\ensuremath{ \wstar_{_{k-2}}  }}
\newcommand{\wstarstar}{\ensuremath{ \wstar^{\star}  }}
\newcommand{\wstarstartwk}{\ensuremath{ \wstarstar_{_{k-2}}  }}

\newcommand{\wezz}{\ensuremath{{ \widehat{\bf w}_{_{0/0} }}}}
\newcommand{\weoz}{\ensuremath{{ \widehat{\bf w}_{_{1/0} }}}}
\newcommand{\weoo}{\ensuremath{{ \widehat{\bf w}_{_{1/1} }}}}
\newcommand{\wekk}{\ensuremath{{ \,\widehat{\bf w}_{_{k/k} }}}}
\newcommand{\wekok}{\ensuremath{{ \,\widehat{\bf w}_{_{k+1/k} }}}}
\newcommand{\weki}{\ensuremath{{ \,\widehat{\bf w}_{_{k/i} }}}}
\newcommand{\wekoK}{\ensuremath{{ \,\widehat{\bf w}_{_{k+1/k} }}}}
\newcommand{\weKok}{\ensuremath{{ \,\widehat{\bf w}_{_{k/k-1} }}}}
\newcommand{\weokk}{\ensuremath{{ \,\widehat{\bf w}_{_{k-1/k} }}}}

\newcommand{\wtv}{\ensuremath{{\bf \omega}^o}}
\newcommand{\wtm}{\ensuremath{{ \Omega}^o}}
\newcommand{\wm}{\ensuremath{{\Omega}}}
\newcommand{\werm}{\ensuremath{{\mathcal{E}}}}
\newcommand{\wers}{\ensuremath{{\sigma_{\varepsilon}}}}

\newcommand{\xettv}{\ensuremath{ \bf{\xi}_{\widehat{x}} }}
\newcommand{\xv}{\ensuremath{ \,{\bf x}  }}
\newcommand{\xedv}{\ensuremath{\widehat{\xv}^d }}
\newcommand{\xev}{\ensuremath{{\widehat{\xv}}}}
\newcommand{\xk}{\ensuremath{ \xv_{_k}  }}
\newcommand{\xko}{\ensuremath{ \xv_{_{k+1}}  }}
\newcommand{\xz}{\ensuremath{ \xv_{_0}  }}
\newcommand{\xo}{\ensuremath{ \xv_{_1}  }}
\newcommand{\xtw}{\ensuremath{ \xv_{_2}  }}
\newcommand{\xsxtk}{\ensuremath{ \xk^{16} }}
\newcommand{\xsxtko}{\ensuremath{ \xko^{16} }}
\newcommand{\xtenk}{\ensuremath{ \xk^{10} }}
\newcommand{\xtenko}{\ensuremath{ \xko^{10} }}
\newcommand{\xninko}{\ensuremath{ \xko^{9} }}
\newcommand{\xnink}{\ensuremath{ \xk^{9} }}
\newcommand{\xekk}{\ensuremath{  \xev_{_{k/k}} }}
\newcommand{\xekok}{\ensuremath{  \xev_{_{k+1/k}} }}
\newcommand{\xezz}{\ensuremath{  \xev_{_{0/0}} }}
\newcommand{\xekoko}{\ensuremath{  \xev_{_{k+1/k+1}} }}
\newcommand{\xzv}{\ensuremath{ \xv^0 }}

\newcommand{\xtv}{\ensuremath{ \xv_{_t}  }}
\newcommand{\xav}{\ensuremath{ \xv_{_a}  }}

\newcommand{\yv}{\ensuremath{{\bf y}}}
\newcommand{\yzv}{\ensuremath{  \yv^0 }}
\newcommand{\yttv}{\ensuremath{ \bf{\xi}_y }}
\newcommand{\Yv}{\ensuremath{ {\bf y}  }}
\newcommand{\ydv}{\ensuremath{\yv^d }}
\newcommand{\yk}{\ensuremath{ \yv_{_k} }}
\newcommand{\yko}{\ensuremath{ \yv_{_{k+1}} }}
\newcommand{\yz}{\ensuremath{ \yv_{_0} }}
\newcommand{\ytv}{\ensuremath{ \yv_{_t}  }}

\newcommand{\yev}{\ensuremath{ \widehat{\yv} }}

\newcommand{\Zv}{\ensuremath{ {\bf z}  }}
\newcommand{\Zeps}{\ensuremath{ {\bf z}_{_\epsilon}  }}
\newcommand{\zeps}{\ensuremath{ {\bf z}_{_\epsilon}  }}
\newcommand{\Zb}{\ensuremath{ {\bf z}_{_b}  }}
\newcommand{\zv}{\ensuremath{ {\bf z}  }}
\newcommand{\zb}{\ensuremath{ {\bf z}_{_b}  }}
\newcommand{\zk}{\ensuremath{ \zv_{_{k}}  }}
\newcommand{\zko}{\ensuremath{ \zv_{_{k+1}}  }}
\newcommand{\dzko}{\ensuremath{ \delta \zv_{_{k+1}}  }}
\newcommand{\zkot}{\ensuremath{ \zv_{_{k+1}}^{true}  }}
\newcommand{\zer}{\ensuremath{ \widetilde{\zv} }}
\newcommand{\zerk}{\ensuremath{ \zer_{_k} }}
\newcommand{\zz}{\ensuremath{ \zv_{_{0}}  }}
\newcommand{\zo}{\ensuremath{ \zv_{_{1}}  }}
\newcommand{\zN}{\ensuremath{ \zv_{_{N}}  }}
\newcommand{\zbi}{\ensuremath{ {\zb^i} }}
\newcommand{\zbn}{\ensuremath{ {\zb^n} }}
\newcommand{\zev}{\ensuremath{ \widehat{\zv} }}
\newcommand{\zzero}{\ensuremath{ {\zv^{o}}  }}
\newcommand{\ztv}{\ensuremath{ \zv_{_{t}}  }}

\newcommand{\zerov}{\ensuremath{\mathbf{0}}}
\newcommand{\onev}{\ensuremath{\mathbf{1}}}


\newcommand{\ai}{\ensuremath{ \,a_{_i}  }}
\newcommand{\aj}{\ensuremath{ \,a_{_j}  }}
\newcommand{\ako}{\ensuremath{ \,a_{_{k+1}}  }}
\newcommand{\ak}{\ensuremath{ \,a_{_k}  }}
\newcommand{\ao}{\ensuremath{ \,a_{_1}  }}
\newcommand{\atw}{\ensuremath{ \,a_{_2}  }}

\newcommand{\alfa}{\ensuremath{ \alpha  }}
\newcommand{\alfaz}{\ensuremath{ \alpha_{_0}  }}
\newcommand{\alfao}{\ensuremath{ \alpha_{_1}  }}
\newcommand{\alfatw}{\ensuremath{ \alpha_{_2}  }}
\newcommand{\alfak}{\ensuremath{ \alpha_{_k}  }}
\newcommand{\alfaN}{\ensuremath{ \,\alpha_{_N}  }}
\newcommand{\alfako}{\ensuremath{ \alpha_{_{k+1}}  }}
\newcommand{\alfai}{\ensuremath{ \alpha_{_i}  }}
\newcommand{\alfaone}{\ensuremath{ \alpha_{_1}  }}
\newcommand{\alfatwo}{\ensuremath{ \alpha_{_2}  }}
\newcommand{\alfathr}{\ensuremath{ \alpha_{_3}  }}
\newcommand{\alfaij}{\ensuremath{ \alpha_{_{ij}}  }}
\newcommand{\alfaonefor}{\ensuremath{ \alpha_{_{14}}  }}

\newcommand{\bos}{\ensuremath{ \,b_{_1}  }}
\newcommand{\btws}{\ensuremath{ \,b_{_2}  }}
\newcommand{\bths}{\ensuremath{ \,b_{_3}  }}

\newcommand{\betat}{\ensuremath{ \beta_{_t}  }}
\newcommand{\betazer}{\ensuremath{ \beta_{_0}  }}
\newcommand{\betaone}{\ensuremath{ \beta_{_1}  }}
\newcommand{\betatwo}{\ensuremath{ \beta_{_2}  }}
\newcommand{\betathr}{\ensuremath{ \beta_{_3}  }}
\newcommand{\betafor}{\ensuremath{ \beta_{_4}  }}
\newcommand{\betai}{\ensuremath{ \beta_{_{\!i}}  }}

\newcommand{\dfi}{\ensuremath{ \;\delta \phi }}
\newcommand{\dfikk}{\ensuremath{ \dfi_{_{k/k}} }}
\newcommand{\dfie}{\ensuremath{ \widetilde{\delta\phi}  }}
\newcommand{\dfiekk}{\ensuremath{ \dfie_{_{k/k}} }}

\newcommand{\deltaok}{\ensuremath{ \,\delta_{_{k-1}}  }}
\newcommand{\deltatwk}{\ensuremath{ \,\delta_{_{k-2}}  }}
\newcommand{\deltaij}{\ensuremath{ \,\delta_{_{ij}}  }}
\newcommand{\dqes}{\ensuremath{ \,\delta q }}

\newcommand{\eos}{\ensuremath{ \,e_{_1}  }}
\newcommand{\etws}{\ensuremath{ \,e_{_2}  }}
\newcommand{\eths}{\ensuremath{ \,e_{_3}  }}

\newcommand{\etaz}{\ensuremath{ \eta_{_0}  }}
\newcommand{\etak}{\ensuremath{ \eta_{_k}  }}
\newcommand{\etaok}{\ensuremath{ \eta_{_{k-1}}  }}
\newcommand{\etai}{\ensuremath{ \eta_{_i}  }}
\newcommand{\etaoz}{\ensuremath{ \eta_{_{1/0}}  }}
\newcommand{\etaoo}{\ensuremath{ \,\eta_{_{1/1}}  }}
\newcommand{\etae}{\ensuremath{\widehat{\eta}}}
\newcommand{\etakast}{\ensuremath{ \etak^{\ast}  }}
\newcommand{\etaokast}{\ensuremath{ \etaok^{\ast}  }}
\newcommand{\etaoN}{\ensuremath{ \,\eta_{_{N-1}}  }}

\newcommand{\eps}{\ensuremath{ \epsilon  }}
\newcommand{\epso}{\ensuremath{ \epsilon_{_1}  }}
\newcommand{\epstw}{\ensuremath{ \epsilon_{_2}  }}
\newcommand{\epsth}{\ensuremath{ \epsilon_{_3}  }}

\newcommand{\fit}{\ensuremath{{ {\varphi}_{_t}}}}

\newcommand{\fie}{\ensuremath{ \widehat{\varphi} }}
\newcommand{\fiet}{\ensuremath{ \fie_{_{t}} }}

\newcommand{\gamazz}{\ensuremath{ \gamma_{_{0/0}}  }}
\newcommand{\gamaoz}{\ensuremath{ \gamma_{_{1/0}}  }}
\newcommand{\gamaoo}{\ensuremath{ \gamma_{_{1/1}}  }}
\newcommand{\gamaoostar}{\ensuremath{ \star{\gamaoo}  }}
\newcommand{\gamakk}{\ensuremath{ \,\gamma_{_{k/k}}  }}
\newcommand{\gamaokok}{\ensuremath{ \,\gamma_{_{ {k-1}/{k-1} }}  }}
\newcommand{\gamakoK}{\ensuremath{ \,\gamma_{_{ {k+1}/{k} }}  }}
\newcommand{\gamaKok}{\ensuremath{  \,\gamma_{_{    k /{k-1} }}  }}
\newcommand{\gamabar}{\ensuremath{  \,\bar{\gamma}  }}
\newcommand{\gamabarKok}{\ensuremath{  \,\gamabar_{_{    k /{k-1} }}  }}
\newcommand{\gamabarbar}{\ensuremath{  \,\bar{\gamabar}  }}
\newcommand{\gamabarbarKok}{\ensuremath{  \,\gamabarbar_{_{    k /{k-1} }}  }}

\newcommand{\gama}{\ensuremath{ \,\gamma  }}
\newcommand{\gamao}{\ensuremath{ \gama_{_{1}}  }}
\newcommand{\gamatw}{\ensuremath{ \gama_{_{2}}  }}
\newcommand{\gamath}{\ensuremath{ \gama_{_{3}}  }}

\newcommand{\gNN}{\ensuremath{  g_{_{N/N}} }}

\newcommand{\Kkoo}{\ensuremath{  \Kk_{_{11}}  }}
\newcommand{\Kkotw}{\ensuremath{ \Kk_{_{12}}  }}
\newcommand{\Kkoth}{\ensuremath{ \Kk_{_{13}}  }}
\newcommand{\Kkof}{\ensuremath{  \Kk_{_{14}}  }}
\newcommand{\Kktwo}{\ensuremath{ \Kk_{_{21}}  }}
\newcommand{\Kktwtw}{\ensuremath{\Kk_{_{22}}  }}
\newcommand{\Kktwth}{\ensuremath{ \Kk_{_{23}}  }}
\newcommand{\Kktwf}{\ensuremath{ \Kk_{_{24}}  }}
\newcommand{\Kktho}{\ensuremath{  \Kk_{_{31}}  }}
\newcommand{\Kkthtw}{\ensuremath{ \Kk_{_{32}}  }}
\newcommand{\Kkthth}{\ensuremath{  \Kk_{_{33}}  }}
\newcommand{\Kkthf}{\ensuremath{ \Kk_{_{34}}  }}
\newcommand{\Kkfo}{\ensuremath{  \Kk_{_{41}}  }}
\newcommand{\Kkftw}{\ensuremath{ \Kk_{_{42}}  }}
\newcommand{\Kkfth}{\ensuremath{ \Kk_{_{43}}  }}
\newcommand{\Kkff}{\ensuremath{  \Kk_{_{44}}  }}
\newcommand{\Kkij}{\ensuremath{  {\Kk}_{_{ij}}  }}

\newcommand{\Kkooo}{\ensuremath{  \Kko_{_{11}}  }}
\newcommand{\Kkootw}{\ensuremath{ \Kko_{_{12}}  }}
\newcommand{\Kkooth}{\ensuremath{ \Kko_{_{13}}  }}
\newcommand{\Kkoof}{\ensuremath{  \Kko_{_{14}}  }}
\newcommand{\Kkotwo}{\ensuremath{ \Kko_{_{21}}  }}
\newcommand{\Kkotwtw}{\ensuremath{\Kko_{_{22}}  }}
\newcommand{\Kkotwth}{\ensuremath{ \Kko_{_{23}}  }}
\newcommand{\Kkotwf}{\ensuremath{ \Kko_{_{24}}  }}
\newcommand{\Kkotho}{\ensuremath{  \Kko_{_{31}}  }}
\newcommand{\Kkothtw}{\ensuremath{ \Kko_{_{32}}  }}
\newcommand{\Kkothth}{\ensuremath{  \Kko_{_{33}}  }}
\newcommand{\Kkothf}{\ensuremath{ \Kko_{_{34}}  }}
\newcommand{\Kkofo}{\ensuremath{  \Kko_{_{41}}  }}
\newcommand{\Kkoftw}{\ensuremath{ \Kko_{_{42}}  }}
\newcommand{\Kkofth}{\ensuremath{ \Kko_{_{43}}  }}
\newcommand{\Kkoff}{\ensuremath{  \Kko_{_{44}}  }}
\newcommand{\Kkoij}{\ensuremath{  {\Kko}_{_{ij}}  }}

\newcommand{\Kon}{\ensuremath{  K_{_{1}}  }}
\newcommand{\Ktw}{\ensuremath{ K_{_{2}}  }}
\newcommand{\Koo}{\ensuremath{  K_{_{11}}  }}
\newcommand{\Kotw}{\ensuremath{ K_{_{12}}  }}
\newcommand{\Koth}{\ensuremath{ K_{_{13}}  }}
\newcommand{\Kof}{\ensuremath{  K_{_{14}}  }}
\newcommand{\Ktwo}{\ensuremath{ K_{_{21}}  }}
\newcommand{\Ktwtw}{\ensuremath{  K_{_{22}}  }}
\newcommand{\Ktwth}{\ensuremath{ K_{_{23}}  }}
\newcommand{\Ktwf}{\ensuremath{ K_{_{24}}  }}
\newcommand{\Ktho}{\ensuremath{  K_{_{31}}  }}
\newcommand{\Kthtw}{\ensuremath{ K_{_{32}}  }}
\newcommand{\Kthth}{\ensuremath{  K_{_{33}}  }}
\newcommand{\Kthf}{\ensuremath{ K_{_{34}}  }}
\newcommand{\Kfo}{\ensuremath{  K_{_{41}}  }}
\newcommand{\Kftw}{\ensuremath{ K_{_{42}}  }}
\newcommand{\Kfth}{\ensuremath{ K_{_{43}}  }}
\newcommand{\Kff}{\ensuremath{  K_{_{44}}  }}
\newcommand{\Ksoneone}{\ensuremath{  K_{_{\!11}}  }}
\newcommand{\Ksonetwo}{\ensuremath{ K_{_{\!12}}  }}
\newcommand{\Ksonethr}{\ensuremath{ K_{_{\!13}}  }}
\newcommand{\Ksonefor}{\ensuremath{  K_{_{\!14}}  }}
\newcommand{\Kstwoone}{\ensuremath{ K_{_{\!21}}  }}
\newcommand{\Kstwotwo}{\ensuremath{  K_{_{\!22}}  }}
\newcommand{\Kstwothr}{\ensuremath{ K_{_{\!23}}  }}
\newcommand{\Kstwofor}{\ensuremath{ K_{_{\!24}}  }}
\newcommand{\Ksthrone}{\ensuremath{  K_{_{\!31}}  }}
\newcommand{\Ksthrtwo}{\ensuremath{ K_{_{\!32}}  }}
\newcommand{\Ksthrthr}{\ensuremath{  K_{_{\!33}}  }}
\newcommand{\Ksthrfor}{\ensuremath{ K_{_{\!34}}  }}
\newcommand{\Ksforone}{\ensuremath{  K_{_{\!41}}  }}
\newcommand{\Ksfortwo}{\ensuremath{ K_{_{\!42}}  }}
\newcommand{\Ksforthr}{\ensuremath{ K_{_{\!43}}  }}
\newcommand{\Ksforfor}{\ensuremath{  K_{_{\!44}}  }}

\newcommand{\lbd}{\ensuremath{{\,\lambda}}}
\newcommand{\lbdz}{\ensuremath{ \lambda^{^0}  }}
\newcommand{\lbdo}{\ensuremath{ \lambda_{_1}  }}
\newcommand{\lbdtw}{\ensuremath{ \lambda_{_2}  }}
\newcommand{\lbdth}{\ensuremath{ \lambda_{_3}  }}
\newcommand{\lbdfo}{\ensuremath{ \lambda_{_4}  }}
\newcommand{\lmx}{\ensuremath{\lambda_{max}}}
\newcommand{\lkok}{\ensuremath{ \,\lambda_{_{k+1/k}} }}
\newcommand{\likok}{\ensuremath{ {\lkok^i} }}
\newcommand{\lzkok}{\ensuremath{ {\lkok^0} }}
\newcommand{\lokok}{\ensuremath{ {\lkok^1} }}
\newcommand{\ltwkok}{\ensuremath{ {\lkok^2} }}
\newcommand{\lthkok}{\ensuremath{ {\lkok^3} }}
\newcommand{\lmxk}{\ensuremath{  {\lmx}_{_k} }}
\newcommand{\lbdi}{\ensuremath{ \lambda_{_i}  }}
\newcommand{\lko}{\ensuremath{ \lambda_{_{k+1}} }}
\newcommand{\liko}{\ensuremath{ \lko^i }}
\newcommand{\lzko}{\ensuremath{ \lko^0 }}
\newcommand{\loko}{\ensuremath{ \lko^1 }}
\newcommand{\ltwko}{\ensuremath{ \lko^2 }}
\newcommand{\lthko}{\ensuremath{ \lko^3 }}

\newcommand{\mk}{\ensuremath{ m_{_k}  }}
\newcommand{\mko}{\ensuremath{ m_{_{k+1}}  }}
\newcommand{\dmko}{\ensuremath{ \delta \mko }}
\newcommand{\mz}{\ensuremath{ m_{_0}  }}
\newcommand{\dmz}{\ensuremath{ \delta \mz }}

\newcommand{\msij}{\ensuremath{ m_{_{ij}}  }}
\newcommand{\msonetwo}{\ensuremath{ m_{_{12}}  }}
\newcommand{\msonethr}{\ensuremath{ m_{_{13}}  }}
\newcommand{\msonefor}{\ensuremath{ m_{_{14}}  }}
\newcommand{\mstwothr}{\ensuremath{ m_{_{23}}  }}
\newcommand{\mstwofor}{\ensuremath{ m_{_{24}}  }}
\newcommand{\msthrfor}{\ensuremath{ m_{_{34}}  }}

\newcommand{\muko}{\ensuremath{ \mu_{_{k+1}}  }}
\newcommand{\mukoast}{\ensuremath{ {\muko^{\ast}}  }}
\newcommand{\muz}{\ensuremath{ \mu_{_0}  }}
\newcommand{\muezz}{\ensuremath{ \widehat{\mu}_{_{0/0}}  }}
\newcommand{\muk}{\ensuremath{ \,\mu_{_k}  }}
\newcommand{\mui}{\ensuremath{ \,\mu_{_i}  }}
\newcommand{\muo}{\ensuremath{ \,\mu_{_1}  }}
\newcommand{\mutw}{\ensuremath{ \,\mu_{_2}  }}
\newcommand{\muth}{\ensuremath{ \,\mu_{_3}  }}
\newcommand{\mufo}{\ensuremath{ \,\mu_{_4}  }}

\newcommand{\nk}{\ensuremath{ {n_{_k}} }}

\newcommand{\nuoz}{\ensuremath{ \nu_{_{1/0}}  }}
\newcommand{\nui}{\ensuremath{ \nu_{_i}  }}

\newcommand{\omo}{\ensuremath{ \,\omega_{_{1}}   }}
\newcommand{\omtw}{\ensuremath{ \,\omega_{_{2}}   }}
\newcommand{\omth}{\ensuremath{ \,\omega_{_{3}}   }}

\newcommand{\omot}{\ensuremath{ \,\omega_{_{1}}^{true}   }}
\newcommand{\omtwt}{\ensuremath{ \,\omega_{_{2}}^{true}   }}
\newcommand{\omtht}{\ensuremath{ \,\omega_{_{3}}^{true}   }}

\newcommand{\Pff}{\ensuremath{ P_{_{44}} }}
\newcommand{\Pfsv}{\ensuremath{ P_{_{47}} }}
\newcommand{\Pfni}{\ensuremath{ P_{_{49}} }}
\newcommand{\Psvsv}{\ensuremath{ P_{_{77}} }}
\newcommand{\Psvni}{\ensuremath{ P_{_{79}} }}
\newcommand{\Pnini}{\ensuremath{ P_{_{99}} }}

\newcommand{\pdf}{\ensuremath{ p_\bv \left( \bv,A \right)  }}

\newcommand{\qs}{\ensuremath{ \,\emph{q}  }}
\newcommand{\qsz}{\ensuremath{ \qs^0 }}
\newcommand{\qsk}{\ensuremath{ \qs_{_k} }}
\newcommand{\qsko}{\ensuremath{ \qs_{_{k+1}} }}
\newcommand{\qes}{\ensuremath{ \,\widehat{\qs}  }}
\newcommand{\qesk}{\ensuremath{ \qes_{_{k}} }}
\newcommand{\qesko}{\ensuremath{ \qes_{_{k+1}} }}
\newcommand{\qeskk}{\ensuremath{ \qes_{_{k/k}} }}
\newcommand{\qeskok}{\ensuremath{ \qes_{_{k+1/k}}  }}
\newcommand{\qeskoko}{\ensuremath{ \qes_{_{k+1/k+1}}  }}
\newcommand{\qeskolk}{\ensuremath{ \qes_{_{k+l-1/k}}  }}
\newcommand{\qesklk}{\ensuremath{ \qes_{_{k+l/k}}  }}
\newcommand{\quats}{\ensuremath{\,\emph{q}}}
\newcommand{\qeps}{\ensuremath{ \,q_{_\epsilon}  }}

\newcommand{\rl}{\ensuremath{ \,r^{\lambda}  }}
\newcommand{\rlko}{\ensuremath{ \rl_{_{k+1}}  }}

\newcommand{\ros}{\ensuremath{ \,r_{_1}  }}
\newcommand{\rtws}{\ensuremath{ \,r_{_2}  }}
\newcommand{\rths}{\ensuremath{ \,r_{_3}  }}

\newcommand{\rkos}{\ensuremath{ \, r_{_{k+1}} }}
\newcommand{\rkoso}{\ensuremath{ {\rkos}_{_1} }}
\newcommand{\rkostw}{\ensuremath{{\rkos}_{_2} }}
\newcommand{\rkosth}{\ensuremath{ {\rkos}_{_3} }}

\newcommand{\rooz}{\ensuremath{ \,\rho_{_{1/0}}  }}
\newcommand{\rozo}{\ensuremath{ \,\rho_{_{0/1}}  }}
\newcommand{\roko}{\ensuremath{ \,\rho_{_{k+1}}  }}
\newcommand{\rok}{\ensuremath{ \,\rho_{_{k}}  }}
\newcommand{\rokoast}{\ensuremath{ \roko^{\ast}  }}
\newcommand{\rokast}{\ensuremath{ \rok^{\ast}  }}
\newcommand{\rooo}{\ensuremath{  \,\rho_{_{1/1}}  }}
\newcommand{\roe}{\ensuremath{   \,\widehat{\rho}}}
\newcommand{\rookk}{\ensuremath{ \,\rho_{_{{k-1}/k}}  }}
\newcommand{\rokok}{\ensuremath{ \,\roe_{_{{k+1}/k}}  }}
\newcommand{\rotwkok}{\ensuremath{ \,\rho_{_{{k-2}/{k-1}}}  }}
\newcommand{\rookok}{\ensuremath{ \,\rho_{_{{k-1}/{k-1}}}  }}
\newcommand{\robar}{\ensuremath{  \,\bar{\rho}  }}
\newcommand{\robarokok}{\ensuremath{ \,\robar_{_{{k-1}/{k-1}}}  }}
\newcommand{\robarbar}{\ensuremath{  \,\bar{\robar}  }}
\newcommand{\robarbarokok}{\ensuremath{ \,\robarbar_{_{{k-1}/{k-1}}}  }}

\newcommand{\sigeps}{\ensuremath{ \sigma_{_{\!\!\!\epsilon}}   }}
\newcommand{\sigb}{\ensuremath{ \,\sigma_{_{\!\!b}}   }}
\newcommand{\kape}{\ensuremath{ \widehat{\kappa} }}

\newcommand{\sigk}{\ensuremath{ \,\sigma_{_{k}}   }}
\newcommand{\sigko}{\ensuremath{ \,\sigma_{_{k+1}}   }}
\newcommand{\dsigko}{\ensuremath{ \,\delta \sigma_{_{k+1}}   }}
\newcommand{\sigkot}{\ensuremath{ \sigko^{true} }}
\newcommand{\sigi}{\ensuremath{ \,\sigma_{_i}   }}
\newcommand{\sigtot}{\ensuremath{ \,\sigma_{_{tot}}   }}
\newcommand{\sigm}{\ensuremath{   \,\sigma_{_m}   }}
\newcommand{\sigbi}{\ensuremath{ {\sigb^i} }}
\newcommand{\sigbn}{\ensuremath{ {\sigb^n} }}
\newcommand{\sigz}{\ensuremath{ \,\sigma^{o} }}
\newcommand{\sig}{\ensuremath{ \sigma }}
\newcommand{\sige}{\ensuremath{ \widehat{\sigma} }}
\newcommand{\sigone}{\ensuremath{ \sig_{_{1}}   }}
\newcommand{\sigtwo}{\ensuremath{ \sig_{_{2}}   }}
\newcommand{\sigthr}{\ensuremath{ \sig_{_{3}}   }}

\newcommand{\tz}{\ensuremath{ \,t_{_0}   }}
\newcommand{\tone}{\ensuremath{ \,t_{_1}   }}
\newcommand{\tk}{\ensuremath{ \,t_{_k}   }}
\newcommand{\tko}{\ensuremath{ \,t_{_{k+1}}   }}
\newcommand{\tktw}{\ensuremath{ \,t_{_{k+2}}   }}
\newcommand{\tok}{\ensuremath{ \,t_{_{k-1}}   }}
\newcommand{\tN}{\ensuremath{ \,t_{_N}   }}
\newcommand{\toN}{\ensuremath{ \,t_{_{N-1}}   }}
\newcommand{\tkN}{\ensuremath{ \,t_{_{k+N}}   }}
\newcommand{\tNo}{\ensuremath{ \,t_{_{N+1}}   }}
\newcommand{\ti}{\ensuremath{ \,t_{_{i}}   }}
\newcommand{\tj}{\ensuremath{ \,t_{_{j}}   }}
\newcommand{\tkl}{\ensuremath{ \,t_{_{k+l}}   }}
\newcommand{\tnu}{\ensuremath{ \,t_{_\nu}   }}
\newcommand{\tro}{\ensuremath{ \,t_{_\rho}   }}

\newcommand{\tetaek}{\ensuremath{ \,\widehat{\Theta}_{_k} }}

\newcommand{\usone}{\ensuremath{  u_{_{1}}  }}
\newcommand{\ustwo}{\ensuremath{  u_{_{2}}  }}
\newcommand{\usthr}{\ensuremath{  u_{_{3}}  }}
\newcommand{\usfor}{\ensuremath{  u_{_{4}}  }}

\newcommand{\vsone}{\ensuremath{  v_{_{1}}  }}
\newcommand{\vstwo}{\ensuremath{  v_{_{2}}  }}
\newcommand{\vsthr}{\ensuremath{  v_{_{3}}  }}
\newcommand{\vsfor}{\ensuremath{  v_{_{4}}  }}

\newcommand{\Vkoioo}{\ensuremath{  \Vkoi_{_{11}}  }}
\newcommand{\Vkoiotw}{\ensuremath{ \Vkoi_{_{12}}  }}
\newcommand{\Vkoioth}{\ensuremath{ \Vkoi_{_{13}}  }}
\newcommand{\Vkoiof}{\ensuremath{  \Vkoi_{_{14}}  }}
\newcommand{\Vkoitwo}{\ensuremath{ \Vkoi_{_{21}}  }}
\newcommand{\Vkoitwtw}{\ensuremath{\Vkoi_{_{22}}  }}
\newcommand{\Vkoitwth}{\ensuremath{ \Vkoi_{_{23}}  }}
\newcommand{\Vkoitwf}{\ensuremath{ \Vkoi_{_{24}}  }}
\newcommand{\Vkoitho}{\ensuremath{  \Vkoi_{_{31}}  }}
\newcommand{\Vkoithtw}{\ensuremath{ \Vkoi_{_{32}}  }}
\newcommand{\Vkoithth}{\ensuremath{  \Vkoi_{_{33}}  }}
\newcommand{\Vkoithf}{\ensuremath{ \Vkoi_{_{34}}  }}
\newcommand{\Vkoifo}{\ensuremath{  \Vkoi_{_{41}}  }}
\newcommand{\Vkoiftw}{\ensuremath{ \Vkoi_{_{42}}  }}
\newcommand{\Vkoifth}{\ensuremath{ \Vkoi_{_{43}}  }}
\newcommand{\Vkoiff}{\ensuremath{  \Vkoi_{_{44}}  }}
\newcommand{\Vkoiij}{\ensuremath{  {\Vkoi}_{_{ij}}  }}

\newcommand{\Wkoo}{\ensuremath{  \Wk_{_{11}}  }}
\newcommand{\Wkotw}{\ensuremath{ \Wk_{_{12}}  }}
\newcommand{\Wkoth}{\ensuremath{ \Wk_{_{13}}  }}
\newcommand{\Wkof}{\ensuremath{  \Wk_{_{14}}  }}
\newcommand{\Wktwo}{\ensuremath{ \Wk_{_{21}}  }}
\newcommand{\Wktwtw}{\ensuremath{\Wk_{_{22}}  }}
\newcommand{\Wktwth}{\ensuremath{ \Wk_{_{23}}  }}
\newcommand{\Wktwf}{\ensuremath{ \Wk_{_{24}}  }}
\newcommand{\Wktho}{\ensuremath{  \Wk_{_{31}}  }}
\newcommand{\Wkthtw}{\ensuremath{ \Wk_{_{32}}  }}
\newcommand{\Wkthth}{\ensuremath{  \Wk_{_{33}}  }}
\newcommand{\Wkthf}{\ensuremath{ \Wk_{_{34}}  }}
\newcommand{\Wkfo}{\ensuremath{  \Wk_{_{41}}  }}
\newcommand{\Wkftw}{\ensuremath{ \Wk_{_{42}}  }}
\newcommand{\Wkfth}{\ensuremath{ \Wk_{_{43}}  }}
\newcommand{\Wkff}{\ensuremath{  \Wk_{_{44}}  }}
\newcommand{\Wkij}{\ensuremath{  {\Wk}_{_{ij}}  }}

\newcommand{\xoo}{\ensuremath{ \,x_{_{11}}   }}
\newcommand{\xotw}{\ensuremath{ \,x_{_{12}}   }}
\newcommand{\xtwtw}{\ensuremath{ \,x_{_{22}}   }}
\newcommand{\xij}{\ensuremath{ \,x_{_{ij}}   }}
\newcommand{\xks}{\ensuremath{ \,x_{_k} }}
\newcommand{\xkso}{\ensuremath{ {\xks}_{_1} }}
\newcommand{\xkstw}{\ensuremath{ {\xks}_{_2} }}
\newcommand{\xksth}{\ensuremath{ {\xks}_{_3} }}
\newcommand{\xksf}{\ensuremath{ {\xks}_{_4} }}
\newcommand{\xksfv}{\ensuremath{ {\xks}_{_5} }}
\newcommand{\xkssx}{\ensuremath{ {\xks}_{_6} }}
\newcommand{\xkssv}{\ensuremath{ {\xks}_{_7} }}
\newcommand{\xksei}{\ensuremath{ {\xks}_{_8} }}
\newcommand{\xksni}{\ensuremath{ {\xks}_{_9} }}

\newcommand{\xst}{\ensuremath{ x_{_{t}}   }}

\newcommand{\xsone}{\ensuremath{ x_{_{1}}   }}
\newcommand{\xstwo}{\ensuremath{ x_{_{2}}   }}
\newcommand{\xsthr}{\ensuremath{ x_{_{3}}   }}
\newcommand{\xsfor}{\ensuremath{ x_{_{4}}   }}

\newcommand{\xes}{\ensuremath{ \,\widehat{x} }}
\newcommand{\xest}{\ensuremath{ \xes_{_{t}}   }}
\newcommand{\xeso}{\ensuremath{ {\xes}_{_1} }}
\newcommand{\xestw}{\ensuremath{ {\xes}_{_2} }}
\newcommand{\xesth}{\ensuremath{ {\xes}_{_3} }}
\newcommand{\xesf}{\ensuremath{ {\xes}_{_4} }}
\newcommand{\xesfv}{\ensuremath{ {\xes}_{_5} }}
\newcommand{\xessx}{\ensuremath{ {\xes}_{_6} }}
\newcommand{\xessv}{\ensuremath{ {\xes}_{_7} }}
\newcommand{\xesei}{\ensuremath{ {\xes}_{_8} }}
\newcommand{\xesni}{\ensuremath{ {\xes}_{_9} }}
\newcommand{\xesten}{\ensuremath{ {\xes}_{_{10}} }}

\newcommand{\xso}{\ensuremath{ {x}_{_1} }}
\newcommand{\xstw}{\ensuremath{ {x}_{_2} }}
\newcommand{\xsth}{\ensuremath{ {x}_{_3} }}
\newcommand{\xsf}{\ensuremath{ {x}_{_4} }}
\newcommand{\xsfv}{\ensuremath{ {x}_{_5} }}
\newcommand{\xssx}{\ensuremath{ {x}_{_6} }}
\newcommand{\xssv}{\ensuremath{ {x}_{_7} }}
\newcommand{\xsei}{\ensuremath{ {x}_{_8} }}
\newcommand{\xsni}{\ensuremath{ {x}_{_9} }}
\newcommand{\xsten}{\ensuremath{ {x}_{_{10}} }}

\newcommand{\xsoneone}{\ensuremath{ {x}_{_{11}} }}
\newcommand{\xstwotwo}{\ensuremath{ {x}_{_{22}} }}
\newcommand{\xsthrthr}{\ensuremath{ {x}_{_{33}} }}
\newcommand{\xsforfor}{\ensuremath{ {x}_{_{44}} }}
\newcommand{\xsonefor}{\ensuremath{ {x}_{_{14}} }}
\newcommand{\xstwofor}{\ensuremath{ {x}_{_{24}} }}
\newcommand{\xsthrfor}{\ensuremath{ {x}_{_{34}} }}
\newcommand{\xsonetwo}{\ensuremath{ {x}_{_{12}} }}
\newcommand{\xsonethr}{\ensuremath{ {x}_{_{13}} }}
\newcommand{\xstwothr}{\ensuremath{ {x}_{_{23}} }}

\newcommand{\xsii}{\ensuremath{ {x}_{_{ii}} }}
\newcommand{\xsjj}{\ensuremath{ {x}_{_{jj}} }}
\newcommand{\xsij}{\ensuremath{ {x}_{_{ij}} }}

\newcommand{\fx}{\ensuremath{ f \left( \xv \right)  }}


\newcommand{\dhat}{\ensuremath{ \,\widehat{D}  }}
\newcommand{\dhato}{\ensuremath{ \dhat_{_1}  }}

\newcommand{\fz}{\ensuremath{ \,F_{_0}  }}
\newcommand{\fN}{\ensuremath{ \,F_{_N}  }}
\newcommand{\foN}{\ensuremath{ \,F_{_{N-1}}  }}
\newcommand{\fk}{\ensuremath{ \,F_{_k}  }}
\newcommand{\foktwk}{\ensuremath{ \,F_{_{k-1,k-2}}  }}
\newcommand{\fKok}{\ensuremath{ \,F_{_{k,k-1}}  }}

\newcommand{\fNqN}{\ensuremath{ \fN \left( \qN \right) }}
\newcommand{\fkqk}{\ensuremath{ \fk \left( \qk \right) }}
\newcommand{\foNqoN}{\ensuremath{ \foN \left( \qoN \right) }}
\newcommand{\fzqz}{\ensuremath{ \fz \left( \qz\right) }}

\newcommand{\fstar}{\ensuremath{ \,{F}^{\star}  }}
\newcommand{\fstarz}{\ensuremath{ \fstar_{_0}  }}
\newcommand{\fstarzqz}{\ensuremath{ \fstarz \left( \qz \right)  }}
\newcommand{\fstarok}{\ensuremath{ \fstar_{_{k-1}}  }}
\newcommand{\fstarokqok}{\ensuremath{ \fstarok \left( \qok \right)  }}
\newcommand{\fstark}{\ensuremath{ \fstar_{_k}  }}
\newcommand{\fstarkqk}{\ensuremath{ \fstark \left( \qk \right)  }}
\newcommand{\fstaroktwk}{\ensuremath{ \fstar_{_{k-1,k-2}}  }}

\newcommand{\fhat}{\ensuremath{ \,\widehat{F}  }}
\newcommand{\fhatz}{\ensuremath{ \fhat_{_0}  }}
\newcommand{\fhatok}{\ensuremath{ \fhat_{_{k-1}}  }}
\newcommand{\fhatk}{\ensuremath{ \fhat_{_k}  }}
\newcommand{\fhatoktwk}{\ensuremath{ \fhat_{_{k-1,k-2}}  }}

\newcommand{\fq}{\ensuremath{ f(\qv) }}

\newcommand{\gz}{\ensuremath{ \,G_{_0}  }}
\newcommand{\gN}{\ensuremath{ \,G_{_N}  }}
\newcommand{\goN}{\ensuremath{ \,G_{_{N-1}}  }}
\newcommand{\gk}{\ensuremath{ \,G_{_k}  }}
\newcommand{\gkqk}{\ensuremath{ \gk \left( \qk \right)  }}
\newcommand{\goktwk}{\ensuremath{ \,G_{_{k-1,k-2}}  }}
\newcommand{\gokok}{\ensuremath{ \,G_{_{k-1,k-1}}  }}
\newcommand{\gNqN}{\ensuremath{ \gN \left( \qN \right) }}

\newcommand{\gstar}{\ensuremath{ \,{G}^{\star}  }}
\newcommand{\gstark}{\ensuremath{ \gstar_{_k}  }}
\newcommand{\gstarkqk}{\ensuremath{ \gstark \left( \qk \right)  }}

\newcommand{\ghat}{\ensuremath{ \,\widehat{G}  }}
\newcommand{\ghatk}{\ensuremath{ \ghat_{_k}  }}

\newcommand{\hN}{\ensuremath{ \,H_{_N}  }}
\newcommand{\hk}{\ensuremath{ \,H_{_k}  }}
\newcommand{\hokok}{\ensuremath{ \,H_{_{k-1,k-1}}  }}
\newcommand{\hKok}{\ensuremath{ \,H_{_{k,k-1}}  }}

\newcommand{\hstar}{\ensuremath{ \,\star{H}  }}
\newcommand{\hstarokok}{\ensuremath{ \hstar_{_{k-1,k-1}}  }}
\newcommand{\hstarKok}{\ensuremath{ \hstar_{_{k,k-1}}  }}

\newcommand{\hhat}{\ensuremath{ \,\widehat{H}  }}
\newcommand{\hhatz}{\ensuremath{ \hhat_{_0}  }}
\newcommand{\hhato}{\ensuremath{ \hhat_{_1}  }}
\newcommand{\hhatokok}{\ensuremath{ \hhat_{_{k-1,k-1}}  }}
\newcommand{\hhatKok}{\ensuremath{ \hhat_{_{k,k-1}}  }}

\newcommand{\iN}{\ensuremath{ \,I_{_N}  }}
\newcommand{\iKok}{\ensuremath{ \,I_{_{k,k-1}}  }}

\newcommand{\istar}{\ensuremath{ \,\star{I}  }}
\newcommand{\istarKok}{\ensuremath{ \istar_{_{k,k-1}}  }}

\newcommand{\ihat}{\ensuremath{ \,\widehat{I}  }}
\newcommand{\ihatKok}{\ensuremath{ \ihat_{_{k,k-1}}  }}

\newcommand{\Jko}{\ensuremath{ \,J_{_{k+1}}  }}
\newcommand{\Jzz}{\ensuremath{ \,J_{_{0/0}}  }}
\newcommand{\Joz}{\ensuremath{ \,J_{_{1/0}}  }}
\newcommand{\Joo}{\ensuremath{ \,J_{_{1/1}}  }}
\newcommand{\Je}{\ensuremath{  \,\widehat{J}}}
\newcommand{\Jezz}{\ensuremath{ \,\Je_{_{0/0}}  }}
\newcommand{\Jeoz}{\ensuremath{ \,\Je_{_{1/0}}  }}
\newcommand{\Jeoo}{\ensuremath{ \,\Je_{_{1/1}}  }}
\newcommand{\Jekk}{\ensuremath{ \,\Je_{_{k/k}}  }}
\newcommand{\Jeokok}{\ensuremath{ \,\Je_{_{k-1/k-1}}  }}
\newcommand{\Jstar}{\ensuremath{  \,\star{J}}}
\newcommand{\Jzzstar}{\ensuremath{ \,\Jstar_{_{0/0}}   }}
\newcommand{\Jozstar}{\ensuremath{ \,\Jstar_{_{1/0}}   }}
\newcommand{\Joostar}{\ensuremath{ \,\Jstar_{_{1/1}}   }}
\newcommand{\Jbar}{\ensuremath{  {  \,\bar{J}} }}
\newcommand{\Joobar}{\ensuremath{  \,\Jbar_{{1/1}} }}
\newcommand{\Joobarstar}{\ensuremath{ \,\Joobar^\star    }}
\newcommand{\Jeoobar}{\ensuremath{  \,\widehat{\Joobar} }}
\newcommand{\JNN}{\ensuremath{ \,J_{_{N/N}}  }}
\newcommand{\JN}{\ensuremath{ \,J_{_{N}}  }}
\newcommand{\JNoN}{\ensuremath{ J_{_{N+1/N}}  }}
\newcommand{\JNoNo}{\ensuremath{ J_{_{N+1/N+1}}  }}


\newcommand{\Aml}{\ensuremath{{ \,\widehat{A}{_{_{ML}}} }}}
\newcommand{\At}{\ensuremath{{ A_{_t}} }}
\newcommand{\Ae}{\ensuremath{  \widehat{A} }}
\newcommand{\Aet}{\ensuremath{  \Ae \left( t \right) }}
\newcommand{\Aetp}{\ensuremath{   \Ae \left( t' \right) }}
\newcommand{\Ako}{\ensuremath{ \,A_{_{k+1}}  }}
\newcommand{\Atp}{\ensuremath{{ \,A \left( t' \right)  }}}
\newcommand{\Atrue}{\ensuremath{ \,A^{\mbox{true}}  }}

\newcommand{\AAm}{\ensuremath{{ \mathcal{A}} }}

\newcommand{\Be}{\ensuremath{ \widehat{B} }}
\newcommand{\Beps}{\ensuremath{{ \,B_{_\epsilon} }}}
\newcommand{\Bb}{\ensuremath{{ \,B_{_b} }}}
\newcommand{\Bt}{\ensuremath{{ \,B \left( t \right)  }}}
\newcommand{\Btp}{\ensuremath{{ \,B \left( t' \right)  }}}
\newcommand{\Btx}{\ensuremath{{ \,B \left( t , \xv\right)  }}}
\newcommand{\Bk}{\ensuremath{{ B_{_{k}}  }}}
\newcommand{\Bko}{\ensuremath{{ \,B_{_{k+1}}  }}}
\newcommand{\dBko}{\ensuremath{{ \,\delta B_{_{k+1}}  }}}
\newcommand{\Bkk}{\ensuremath{{ \,B_{_{k/k}}  }}}
\newcommand{\Bkok}{\ensuremath{{ \,B_{_{k+1/k}}  }}}
\newcommand{\Bkoko}{\ensuremath{{ \,B_{_{k+1/k+1}}  }}}
\newcommand{\bicross}{\ensuremath{ \left[ {\bf b}_{_i} \times \right]  }}
\newcommand{\bcross}{\ensuremath{ \left[ {\bf b} \times \right]  }}
\newcommand{\bkcross}{\ensuremath{ \left[ {\bk} \times \right]  }}

\newcommand{\Bbi}{\ensuremath{ \Bb^i }}
\newcommand{\Bbn}{\ensuremath{ \Bb^n }}
\newcommand{\Bz}{\ensuremath{{ \,B^o }}}

\newcommand{\Btrue}{\ensuremath{ \,B^o }}
\newcommand{\Bkkt}{\ensuremath{ \Btrue_{_{k/k}} }}
\newcommand{\Bkot}{\ensuremath{ \Btrue_{_{k+1}} }}

\newcommand{\Coo}{\ensuremath{{ \,C{_{_{11}}} }}}
\newcommand{\Cotw}{\ensuremath{{ \,C{_{_{12}}} }}}
\newcommand{\Ctwo}{\ensuremath{{ \,C{_{_{21}}} }}}
\newcommand{\Ctwtw}{\ensuremath{{ \,C{_{_{22}}} }}}

\newcommand{\Cbar}{\ensuremath{\overline{C} }}
\newcommand{\Cbark}{\ensuremath{ \Cbar_{_{k}} }}

\newcommand{\Cl}{\ensuremath{  \,\mbox{Cl} }}
\newcommand{\Clo}{\ensuremath{  \Cl_{_1} }}
\newcommand{\Cltw}{\ensuremath{  \Cl_{_2} }}
\newcommand{\Clth}{\ensuremath{  \Cl_{_3} }}
\newcommand{\Clf}{\ensuremath{  \Cl_{_4} }}
\newcommand{\Clfv}{\ensuremath{  \Cl_{_5} }}
\newcommand{\Clsx}{\ensuremath{  \Cl_{_6} }}
\newcommand{\Clsv}{\ensuremath{  \Cl_{_7} }}
\newcommand{\Clei}{\ensuremath{  \Cl_{_8} }}
\newcommand{\Clni}{\ensuremath{  \Cl_{_9} }}
\newcommand{\Clten}{\ensuremath{  \Cl_{_{10}} }}
\newcommand{\Clel}{\ensuremath{  \Cl_{_{11}} }}
\newcommand{\Cltwl}{\ensuremath{  \Cl_{_{12}} }}
\newcommand{\Clthr}{\ensuremath{  \Cl_{_{13}} }}
\newcommand{\Clfrt}{\ensuremath{  \Cl_{_{14}} }}
\newcommand{\Clfft}{\ensuremath{  \Cl_{_{15}} }}
\newcommand{\Clsxt}{\ensuremath{  \Cl_{_{16}} }}

\newcommand{\Dtw}{\ensuremath{  \,D_{_2} }}

\newcommand{\Deltay}{\ensuremath{  \Delta_{_{\!Y}} }}

\newcommand{\Di}{\ensuremath{ D_{_{i}} }}
\newcommand{\Dt}{\ensuremath{{ D_{_t}} }}

\newcommand{\dbicross}{\ensuremath{ \left[ \delta {\bf b}_{_i} \times \right]  }}
\newcommand{\dbjcross}{\ensuremath{ \left[ \delta {\bf b}_{_j} \times \right]  }}
\newcommand{\dbkocross}{\ensuremath{ \left[ \dbko \times \right]  }}
\newcommand{\donevcross}{\ensuremath{ \left[ \donev \times \right]  }}

\newcommand{\Eij}{\ensuremath{{ \, \Xer{_{_{ij}}} }}}
\newcommand{\Eji}{\ensuremath{{ \,\Xer{_{_{ji}}} }}}
\newcommand{\Eik}{\ensuremath{{ \,\Xer{_{_{ik}}} }}}
\newcommand{\Ejk}{\ensuremath{{ \,\Xer{_{_{jk}}} }}}
\newcommand{\Ekk}{\ensuremath{{ \,\Xer{_{_{k/k}}} }}}
\newcommand{\Ekok}{\ensuremath{{ \,\Xer{_{_{k+1/k}}} }}}
\newcommand{\Ekoko}{\ensuremath{{ \,\Xer{_{_{k+1/k+1}}} }}}
\newcommand{\ekcross}{\ensuremath{ \,\left[ \ek \times \right] }}
\newcommand{\ekocross}{\ensuremath{ \,\left[ \eko \times \right] }}
\newcommand{\ecross}{\ensuremath{ \,\left[ \ev \times \right] }}
\newcommand{\evcross}{\ensuremath{ \,\left[ \ev \times \right] }}
\newcommand{\eekcross}{\ensuremath{ \,\left[ \eek \times \right] }}
\newcommand{\eekkcross}{\ensuremath{ \,\left[ \eekk \times \right] }}
\newcommand{\eekocross}{\ensuremath{ \,\left[ \eeko \times \right] }}
\newcommand{\eekokcross}{\ensuremath{ \,\left[ \eekok \times \right] }}
\newcommand{\eekokocross}{\ensuremath{ \,\left[ \eekoko \times \right] }}
\newcommand{\eekolkcross}{\ensuremath{ \,\left[ \eekolk \times \right] }}
\newcommand{\eeklkcross}{\ensuremath{ \,\left[ \eeklk \times \right] }}

\newcommand{\Epsk}{\ensuremath{{ \,\mathcal{E}_{_k}}}}
\newcommand{\Epsko}{\ensuremath{{ \,\mathcal{E}_{_{k+1}}}}}
\newcommand{\Eps}{\ensuremath{ \,\mathcal{E} }}
\newcommand{\epscross}{\ensuremath{ \,\left[ \epsv \times \right] }}
\newcommand{\epskocross}{\ensuremath{ \,\left[ \epsko \times \right] }}
\newcommand{\epskcross}{\ensuremath{ \,\left[ \epsk \times \right] }}

\newcommand{\Fko}{\ensuremath{ \,F_{_{k+1}} }}

\newcommand{\Ftt}{\ensuremath{ \,F_{_{\theta \theta}}  }}
\newcommand{\Fk}{\ensuremath{ \,\mathcal{F}_{_k}  }}

\newcommand{\Fik}{\ensuremath{   \,\Phi_{_k} }}
\newcommand{\FiN}{\ensuremath{   \,\Phi_{_N} }}
\newcommand{\Fiko}{\ensuremath{   \,\Phi_{_{k+1}} }}
\newcommand{\Fikt}{\ensuremath{   {\Phi_{_k}^{o}} }}
\newcommand{\Fiqk}{\ensuremath{   \,\Phi_{4_k} }}
\newcommand{\Fiz}{\ensuremath{   \,\Phi_{_0} }}
\newcommand{\dFik}{\ensuremath{  \,\Delta \Phi_{_k} }}
\newcommand{\Fitt}{\ensuremath{  \, \Phi \left( t' , t \right)   }}
\newcommand{\fikcross}{\ensuremath{ \,\left[ \fik \times \right] }}
\newcommand{\Fikol}{\ensuremath{ \, \Phi_{_{k+l-1}} }}
\newcommand{\Fiok}{\ensuremath{{   \,\Phi_{_{k-1}}  }}}
\newcommand{\FioN}{\ensuremath{{   \,\Phi_{_{N-1}}  }}}
\newcommand{\Fiktrue}{\ensuremath{{  \,\Phi^0_{_k}}}}
\newcommand{\Fie}{\ensuremath{   \,\widehat{\Phi} }}
\newcommand{\Fiekk}{\ensuremath{  \Fie_{_{k/k}}  }}
\newcommand{\Fieokok}{\ensuremath{  \Fie_{_{k-1/k-1}}  }}
\newcommand{\Fiezz}{\ensuremath{  \Fie_{_{0/0}}  }}

\newcommand{\Fito}{\ensuremath{ F_{_{\!\!I}} }}
\newcommand{\Flgv}{\ensuremath{ F_{_{\!\!L}} }}
\newcommand{\Fy}{\ensuremath{ F_{_{\!\!Y}} }}

\newcommand{\Fisxtk}{\ensuremath{   \Fik^{16} }}
\newcommand{\Fitenk}{\ensuremath{   \Fik^{10} }}
\newcommand{\Finink}{\ensuremath{   \Fik^{9} }}

\newcommand{\fitm}{\ensuremath{{ \Phi}^o}}
\newcommand{\fim}{\ensuremath{{ \Phi}}}
\newcommand{\fierm}{\ensuremath{{\Delta \Phi}}}
\newcommand{\fikrm}{\ensuremath{\fim_{[2]_k} }}

\newcommand{\Gam}{\ensuremath{ \,\Gamma }}
\newcommand{\Gamk}{\ensuremath{ \Gam_{_{k}} }}
\newcommand{\Gamae}{\ensuremath{ \widehat{\Gam} }}
\newcommand{\Gamekk}{\ensuremath{ \Gamae_{_{k/k}} }}

\newcommand{\Gamsxtk}{\ensuremath{ \Gamk^{16} }}
\newcommand{\Gamtenk}{\ensuremath{ \Gamk^{10} }}
\newcommand{\Gamnink}{\ensuremath{ \Gamk^{9} }}

\newcommand{\Gko}{\ensuremath{ \,G_{_{k+1}} }}
\newcommand{\GWLS}{\ensuremath{ \,G^{_{_{WLS}}} }}
\newcommand{\GWLSko}{\ensuremath{ \GWLS_{_{k+1}} }}

\newcommand{\Hk}{\ensuremath{ \,H_{_{k}} }}
\newcommand{\Hko}{\ensuremath{ \,H_{_{k+1}} }}
\newcommand{\Hzz}{\ensuremath{ \,H^{0}_{_{0}} }}
\newcommand{\Hzo}{\ensuremath{ \,H^{1}_{_{0}} }}
\newcommand{\Ho}{\ensuremath{ \,H_{_{1}} }}
\newcommand{\HN}{\ensuremath{ \,H_{_{N}} }}
\newcommand{\HNo}{\ensuremath{ \,H_{_{N+1}} }}
\newcommand{\Hktw}{\ensuremath{ \,H_{_{k+2}} }}
\newcommand{\Hkl}{\ensuremath{ \,H_{_{k+l}} }}
\newcommand{\Hkot}{\ensuremath{ \Hko^{true} }}
\newcommand{\Herko}{\ensuremath{ \,\Delta H_{_{k+1}} }}
\newcommand{\Hglobal}{\ensuremath{ \,\mathcal{H} }}
\newcommand{\Hglobaloo}{\ensuremath{ \Hglobal_{_{1/1}} }}
\newcommand{\HglobalNN}{\ensuremath{ \Hglobal_{_{N/N}} }}
\newcommand{\HglobalNoN}{\ensuremath{ \Hglobal_{_{N+1/N}} }}
\newcommand{\HglobalNoNo}{\ensuremath{ \Hglobal_{_{N+1/N+1}} }}
\newcommand{\Hglobalkk}{\ensuremath{ \Hglobal_{_{k/k}} }}
\newcommand{\Hglobalkok}{\ensuremath{ \Hglobal_{_{k+1/k}} }}
\newcommand{\Hglobalkoko}{\ensuremath{ \Hglobal_{_{k+1/k+1}} }}
\newcommand{\Hglobalzz}{\ensuremath{ \Hglobal_{_{0/0}} }}

\newcommand{\Hbar}{\ensuremath{\overline{H} }}
\newcommand{\Hbark}{\ensuremath{ \Hbar_{_{k}} }}
\newcommand{\Hbarko}{\ensuremath{ \Hbar_{_{k+1}} }}

\newcommand{\Itwo}{\ensuremath{ \,I_{_2}  }}
\newcommand{\Ithree}{\ensuremath{ \,I_{_3}  }}
\newcommand{\Ifour}{\ensuremath{  \,I_{_4}  }}
\newcommand{\Inine}{\ensuremath{  \,I_{_9}  }}
\newcommand{\Idm}{\ensuremath{ \,I_{_m}  }}
\newcommand{\Isixteen}{\ensuremath{  \,I_{_{16}}  }}

\newcommand{\Kk}{\ensuremath{{ K_{_k} }}}
\newcommand{\Kko}{\ensuremath{ K_{_{k+1}} }}
\newcommand{\Ky}{\ensuremath{{ K_{_{\!Y}} }}}
\newcommand{\Kepsk}{\ensuremath{{ K_{_{\epsk}} }}}
\newcommand{\Kbko}{\ensuremath{{ K_{_{b_{k+1}}} }}}
\newcommand{\Kbk}{\ensuremath{{ K_{_{b_{k}}} }}}
\newcommand{\Kkk}{\ensuremath{{ \,K{_{_{\!\!k/k}}}  }}}
\newcommand{\Kij}{\ensuremath{{ \,K{_{_{\!\!i/j}}}  }}}
\newcommand{\Kijt}{\ensuremath{{ \,K_{_{\!\!i/j}}^{o}  }}}
\newcommand{\KNN}{\ensuremath{{ \,K{_{_{\!\!N/N}}}  }}}
\newcommand{\Kkok}{\ensuremath{{ \,K_{_{\!\!k+1/k}}  }}}
\newcommand{\Kkoko}{\ensuremath{{ \,K_{_{\!\!k+1/k+1}}  }}}
\newcommand{\Kzz}{\ensuremath{{ \,K{_{_{\!\!0/0}}}  }}}
\newcommand{\dK}{\ensuremath{ \,\delta\!K }}
\newcommand{\dKko}{\ensuremath{ \,\delta\!K_{_{\!k+1}} }}
\newcommand{\dKk}{\ensuremath{ \,\delta\!K_{_{\!k}} }}
\newcommand{\dKi}{\ensuremath{ \,\delta\!K_{_{\!i}} }}
\newcommand{\dKit}{\ensuremath{ \delta\!K_{_{i}}^{o}  }}
\newcommand{\dKz}{\ensuremath{ \,\delta\!K_{_{0}} }}
\newcommand{\dKon}{\ensuremath{ \,\delta\!K_{_{1}} }}
\newcommand{\dKtw}{\ensuremath{ \,\delta\!K_{_{2}} }}
\newcommand{\Kwerm}{\ensuremath{{{{\Delta\!K}^\varepsilon} }}}
\newcommand{\DKkokeps}{\ensuremath{ \Kwerm_{_{k+1/k}} }}
\newcommand{\Kweronem}{\ensuremath{ \Kwerm^{(1)}}}
\newcommand{\Kberm}{\ensuremath{{{{\Delta K}^b} }}}
\newcommand{\Ker}{\ensuremath{  \Delta\!K  }}
\newcommand{\Kzer}{\ensuremath{  {\Ker^0} }}
\newcommand{\Kerij}{\ensuremath{  \Ker_{_{i/j}}  }}
\newcommand{\Kerkk}{\ensuremath{  \Ker_{_{k/k}}  }}
\newcommand{\Kerkok}{\ensuremath{  \Ker_{_{k+1/k}}  }}
\newcommand{\Kzerkok}{\ensuremath{  {\Kerkok^0} }}
\newcommand{\Kzerkoko}{\ensuremath{  {\Kerkoko^0} }}
\newcommand{\Kzerkk}{\ensuremath{  {\Kerkk^0} }}
\newcommand{\Kierkok}{\ensuremath{  {\Kerkok^i} }}
\newcommand{\Kerkoko}{\ensuremath{  \Ker_{_{k+1/k+1}}  }}
\newcommand{\Ke}{\ensuremath{ \,\widehat{K} }}
\newcommand{\Kekk}{\ensuremath{ \Ke_{_{k/k}} }}

\newcommand{\Kt}{\ensuremath{ \,K^o  }}
\newcommand{\Ktij}{\ensuremath{ \Kt_{_{i/j}}  }}
\newcommand{\dKt}{\ensuremath{  \dK^o }}
\newcommand{\dKti}{\ensuremath{ \dKt_{_i}  }}
\newcommand{\dKkot}{\ensuremath{ \dKt_{_{k+1}} }}
\newcommand{\Kzzt}{\ensuremath{ \Kt_{_{0/0}} }}
\newcommand{\Kkokt}{\ensuremath{ \Kt_{_{k+1/k}} }}
\newcommand{\Kkkt}{\ensuremath{ \Kt_{_{k/k}} }}
\newcommand{\Kkokot}{\ensuremath{ \Kt_{_{k+1/k+1}} }}
\newcommand{\Kkokap}{\ensuremath{ \Kkokt }}
\newcommand{\Kkkap}{\ensuremath{ \Kkkt }}
\newcommand{\Kkokoap}{\ensuremath{ \Kkokot }}
\newcommand{\Kzzap}{\ensuremath{ \Kzzt }}

\newcommand{\Kgaint}{\ensuremath{ K_{_{t}}  }}

\newcommand{\KW}{\ensuremath{ \,K^{_{_W} } }}
\newcommand{\KWko}{\ensuremath{ \KW_{_{k+1}} }}

\newcommand{\Lko}{\ensuremath{ \,L_{_{k+1}} }}
\newcommand{\Ltw}{\ensuremath{ \,L_{_{2}} }}
\newcommand{\LN}{\ensuremath{ \,L_{_{N}} }}
\newcommand{\Lt}{\ensuremath{ L_{_{t}} }}
\newcommand{\Lonefor}{\ensuremath{ L_{_{14}} }}
\newcommand{\Ltwofor}{\ensuremath{ L_{_{24}} }}
\newcommand{\Lthrfor}{\ensuremath{ L_{_{34}} }}
\newcommand{\Ljk}{\ensuremath{ L_{_{jk}} }}
\newcommand{\Lij}{\ensuremath{ L_{_{ij}} }}

\newcommand{\Ll}{\ensuremath{  \,\mbox{Li} }}
\newcommand{\Llo}{\ensuremath{  \Ll_{_1} }}
\newcommand{\Lltw}{\ensuremath{  \Ll_{_2} }}
\newcommand{\Llth}{\ensuremath{  \Ll_{_3} }}
\newcommand{\Llf}{\ensuremath{  \Ll_{_4} }}
\newcommand{\Llfv}{\ensuremath{  \Ll_{_5} }}
\newcommand{\Llsx}{\ensuremath{  \Ll_{_6} }}
\newcommand{\Llsv}{\ensuremath{  \Ll_{_7} }}
\newcommand{\Llei}{\ensuremath{  \Ll_{_8} }}
\newcommand{\Llni}{\ensuremath{  \Ll_{_9} }}
\newcommand{\Llten}{\ensuremath{  \Ll_{_{10}} }}
\newcommand{\Llel}{\ensuremath{  \Ll_{_{11}} }}
\newcommand{\Lltwl}{\ensuremath{  \Ll_{_{12}} }}
\newcommand{\Llthr}{\ensuremath{  \Ll_{_{13}} }}
\newcommand{\Llfrt}{\ensuremath{  \Ll_{_{14}} }}
\newcommand{\Llfft}{\ensuremath{  \Ll_{_{15}} }}
\newcommand{\Llsxt}{\ensuremath{  \Ll_{_{16}} }}

\newcommand{\Lam}{\ensuremath{ \,\Lambda }}
\newcommand{\Lamko}{\ensuremath{ \Lam_{_{k+1}} }}
\newcommand{\Lami}{\ensuremath{ \Lam^i }}
\newcommand{\Lamiko}{\ensuremath{ {\Lami_{_{k+1}}} }}
\newcommand{\Lamj}{\ensuremath{ \Lam^j }}
\newcommand{\Lamjko}{\ensuremath{ {\Lamj_{_{k+1}}} }}

\newcommand{\Mk}{\ensuremath{ \,M_{_k} }}
\newcommand{\Mtw}{\ensuremath{ \,M_{_2} }}

\newcommand{\onem}{\ensuremath{\,\widehat{R}}}
\newcommand{\Omg}{\ensuremath{ \,\Omega }}
\newcommand{\Omk}{\ensuremath{ \,\Omega_{_k} }}
\newcommand{\Omkt}{\ensuremath{ \,\Omega_{_k}^{true} }}
\newcommand{\omgcross}{\ensuremath{ \,\left[ \omg \times \right] }}
\newcommand{\omkcross}{\ensuremath{ \,\left[ \omk \times \right] }}
\newcommand{\omktcross}{\ensuremath{ \,\left[ \omkt \times \right] }}
\newcommand{\omekkcross}{\ensuremath{ \,\left[ \omekk \times \right] }}
\newcommand{\Ome}{\ensuremath{ \widehat{\Omg} }}
\newcommand{\Omekk}{\ensuremath{ \Ome_{_{k/k}}  }}

\newcommand{\Omghat}{\ensuremath{ \widehat{\Omega} }}

\newcommand{\Ofour}{\ensuremath{ \,O_{_4} }}
\newcommand{\Othree}{\ensuremath{ \,O_{_3} }}

\newcommand{\Pij}{\ensuremath{{ \,P{_{_{i/j}}} }}}
\newcommand{\Pkk}{\ensuremath{ \,P_{_{k/k}}  }}
\newcommand{\Pkok}{\ensuremath{ \,P_{_{k+1/k}} }}
\newcommand{\Pklk}{\ensuremath{{ \,P_{_{k+l/k}} }}}
\newcommand{\Pkolk}{\ensuremath{{ \,P_{_{k+l-1/k}} }}}
\newcommand{\Pkoko}{\ensuremath{ \,P_{_{k+1/k+1}} }}
\newcommand{\Past}{\ensuremath{ \,P^{\ast} }}
\newcommand{\Pkokoast}{\ensuremath{ \Past_{_{k+1/k+1}} }}
\newcommand{\Pkkast}{\ensuremath{ \Past_{_{k/k}} }}
\newcommand{\Pzz}{\ensuremath{{ \,P_{_{0/0}} }}}
\newcommand{\Ptt}{\ensuremath{ \,P_{_{\theta \theta}}  }}
\newcommand{\Pxx}{\ensuremath{{ \,P{_{_{XX}}} }}}
\newcommand{\Pxxoo}{\ensuremath{ \Pxx_{_{11}}       }}
\newcommand{\Prr}{\ensuremath{{ \,P{_{_{\!\!\!rr}}} }}}
\newcommand{\Prrkok}{\ensuremath{ \Prr_{_{k+1/k}} }}
\newcommand{\Prrklk}{\ensuremath{ \Prr_{_{k+l/k}} }}
\newcommand{\Pmm}{\ensuremath{{ \,P{_{_{mm}}} }}}
\newcommand{\Pko}{\ensuremath{ \,P_{_{k+1}} }}
\newcommand{\Pk}{\ensuremath{ \,P_{_{k}} }}

\newcommand{\Pt}{\ensuremath{ P_{_{t}} }}

\newcommand{\Pq}{\ensuremath{ \,P^{q} }}
\newcommand{\Pqkok}{\ensuremath{ \Pq_{_{\!\!k+1/k}} }}
\newcommand{\Pqkoko}{\ensuremath{ \Pq_{_{\!\!k+1/k+1}} }}
\newcommand{\Pqkk}{\ensuremath{ \Pq_{_{\!\!k/k}} }}

\newcommand{\Pee}{\ensuremath{ P_{_{ee}} }}
\newcommand{\Pzzz}{\ensuremath{ P_{_{zz}} }}

\newcommand{\Pv}{\ensuremath{ P^{v} }}
\newcommand{\Pvk}{\ensuremath{ \Pv_{_{\!\!k}} }}
\newcommand{\Pvij}{\ensuremath{ P^{v_{ij}} }}
\newcommand{\Pvonefor}{\ensuremath{ P^{v_{14}} }}

\newcommand{\Pvbar}{\ensuremath{ P^{\bar{v}} }}
\newcommand{\Pvbark}{\ensuremath{ \Pvbar_{_{\!\!k}} }}
\newcommand{\Pvbarko}{\ensuremath{ \Pvbar_{_{\!\!k+1}} }}

\newcommand{\PDqsca}{\ensuremath{ P_{_{\!\!\!\!\Dqsca}} }}
\newcommand{\PDqbar}{\ensuremath{ P_{_{\!\!\!\!\Dqbar}} }}
\newcommand{\PDqev}{\ensuremath{ P_{_{\!\!\!\!\Dqev}} }}
\newcommand{\Pdelqbar}{\ensuremath{ P_{_{\!\!\!\!\delqbar}} }}
\newcommand{\Pdelqev}{\ensuremath{ P_{_{\!\!\!\!\delqev}} }}
\newcommand{\PDbonev}{\ensuremath{ P_{_{\!\!\!\!\Dbonev}} }}
\newcommand{\PDronev}{\ensuremath{ P_{_{\!\!\!\!\Dronev}} }}
\newcommand{\PDbtwov}{\ensuremath{ P_{_{\!\!\!\!\Dbtwov}} }}
\newcommand{\PDrtwov}{\ensuremath{ P_{_{\!\!\!\!\Drtwov}} }}
\newcommand{\PDdv}{\ensuremath{ P_{_{\!\!\!\!\Ddv}} }}
\newcommand{\PDsv}{\ensuremath{ P_{_{\!\!\!\!\Dsv}} }}
\newcommand{\PDbv}{\ensuremath{ P_{_{\!\!\!\!\Dbv}} }}
\newcommand{\PDrv}{\ensuremath{ P_{_{\!\!\!\!\Drv}} }}
\newcommand{\PDbvDrv}{\ensuremath{ P_{_{\!\!\!\!\Dbv\Drv}} }}
\newcommand{\PDdonevDdtwov}{\ensuremath{ P_{_{\!\!\!\!\Ddonev\!\Ddtwov}} }}
\newcommand{\PDdonevDsonev}{\ensuremath{ P_{_{\!\!\!\!\Ddonev\!\Dsonev}} }}
\newcommand{\PDdtwovDsonev}{\ensuremath{ P_{_{\!\!\!\!\Ddtwov\!\Dsonev}} }}
\newcommand{\PDsonev}{\ensuremath{ P_{_{\!\!\!\!\Dsonev}} }}
\newcommand{\PDsonevDdonev}{\ensuremath{ P_{_{\!\!\!\!\Dsonev\!\Ddonev}} }}
\newcommand{\PDsonevDdtwov}{\ensuremath{ P_{_{\!\!\!\!\Dsonev\!\Ddtwov}} }}
\newcommand{\PDdonev}{\ensuremath{ P_{_{\!\!\!\!\Ddonev}} }}
\newcommand{\PDdtwov}{\ensuremath{ P_{_{\!\!\!\!\Ddtwov}} }}

\newcommand{\Psik}{\ensuremath{ \,\Psi_{_k} }}

\newcommand{\Qk}{\ensuremath{ \,\mathcal{Q}{_{_k}} }}
\newcommand{\Qoo}{\ensuremath{{ \,\mathcal{Q}{_{_{11}}} }}}
\newcommand{\Qot}{\ensuremath{{ \,\mathcal{Q}{_{_{12}}} }}}
\newcommand{\Qto}{\ensuremath{{ \,\mathcal{Q}{_{_{21}}} }}}
\newcommand{\Qtt}{\ensuremath{{ \,\mathcal{Q}{_{_{22}}} }}}
\newcommand{\Qwk}{\ensuremath{  \,Q_{_{w_k}}   }}
\newcommand{\Qeps}{\ensuremath{ Q^{\eps} }}
\newcommand{\Qkeps}{\ensuremath{ \Qeps_{_k} }}
\newcommand{\Qepsk}{\ensuremath{ \Qeps_{_k} }}
\newcommand{\Qkleps}{\ensuremath{ \Qeps_{_{k+l}} }}
\newcommand{\Qq}{\ensuremath{ Q^{q} }}
\newcommand{\Qqk}{\ensuremath{ \Qq_{_{k}} }}
\newcommand{\Qthrfor}{\ensuremath{ Q_{_{\!34}} }}

\newcommand{\Qt}{\ensuremath{ Q{_{_t}} }}

\newcommand{\Rk}{\ensuremath{{ {R}{_{_k}} }}}
\newcommand{\Rko}{\ensuremath{{ {R}{_{_{k+1}}} }}}

\newcommand{\Rz}{\ensuremath{{ \,\mathcal{R}{_{_0}} }}}
\newcommand{\Rb}{\ensuremath{ \,\mathcal{R}^{b}  }}
\newcommand{\Rbi}{\ensuremath{ {\Rb^i} }}
\newcommand{\Rkob}{\ensuremath{ \Rb_{_{k+1}} }}
\newcommand{\Rkobi}{\ensuremath{ {{\Rkob}^i} }}
\newcommand{\Rktwb}{\ensuremath{ \Rb_{_{k+2}} }}
\newcommand{\Rklb}{\ensuremath{ \Rb_{_{k+l}} }}
\newcommand{\Rq}{\ensuremath{ R^{q} }}
\newcommand{\Rqko}{\ensuremath{ \Rq_{_{k+1}} }}
\newcommand{\Rkoq}{\ensuremath{ \Rqko }}
\newcommand{\Roo}{\ensuremath{{ \,\mathcal{R}{_{_{11}}} }}}
\newcommand{\Rot}{\ensuremath{{ \,\mathcal{R}{_{_{12}}} }}}
\newcommand{\Rto}{\ensuremath{{ \,\mathcal{R}{_{_{21}}} }}}
\newcommand{\Rtt}{\ensuremath{{ \,\mathcal{R}{_{_{22}}} }}}
\newcommand{\rkocross}
{\ensuremath{ \,\left[ \rko \times \right] }}

\newcommand{\Rt}{\ensuremath{ R_{_{t}} }}
\newcommand{\Rn}{\ensuremath{ R_{_{n}} }}

\newcommand{\Seps}{\ensuremath{{ \,S_{_\epsilon} }}}
\newcommand{\Sk}{\ensuremath{ \,S_{_{k}} }}
\newcommand{\Sko}{\ensuremath{ \,S_{_{k+1}} }}
\newcommand{\dSko}{\ensuremath{ \,\delta S_{_{k+1}} }}
\newcommand{\Skot}{\ensuremath{ \,S_{_{k+1}}^{true} }}
\newcommand{\Sbb}{\ensuremath{{ \,\mathcal{S}{_{_b}} }}}
\newcommand{\scross}{\ensuremath{ \,\left[ \sv \times \right] }}
\newcommand{\skcross}{\ensuremath{ \,\left[ \sk \times \right] }}
\newcommand{\skocross}{\ensuremath{ \,\left[ \sko \times \right] }}
\newcommand{\Sbbi}{\ensuremath{ {\Sbb^i} }}
\newcommand{\Sbbn}{\ensuremath{ {\Sbb^n} }}
\newcommand{\Sz}{\ensuremath{ \,S^{o} }}

\newcommand{\Tetako}{\ensuremath{ \,\Theta_{_{k+1}}    }}
\newcommand{\Tetakok}{\ensuremath{ \,\Theta_{_{k+1/k}} }}
\newcommand{\Tetae}{\ensuremath{ \widehat{\Theta} }}
\newcommand{\Tetaeko}{\ensuremath{ \Tetae_{_{k+1}} }}

\newcommand{\Ukok}{\ensuremath{{ \,U_{_{k+1/k}} }}}
\newcommand{\Ucovkok}{\ensuremath{ \,\mathcal{U}_{_{k+1/k}}  }}
\newcommand{\ucross}{\ensuremath{ \,\left[ \uv \times \right] }}
\newcommand{\uivcross}{\ensuremath{ \,\left[ \uiv \times \right] }}

\newcommand{\Vk}{\ensuremath{{ \,V_{_k} }}}
\newcommand{\Vi}{\ensuremath{{ \,V_{_i} }}}
\newcommand{\Vko}{\ensuremath{{ \,V_{_{\!\!k+1}} }}}
\newcommand{\Voj}{\ensuremath{{ \,V_{_{1j}} }}}
\newcommand{\Voi}{\ensuremath{{ \,V_{_{1i}} }}}
\newcommand{\vcross}{\ensuremath{ \,\left[ \vvv \times \right] }}
\newcommand{\Vkoi}{\ensuremath{ {\,V_{_{k+1}}^i} }}
\newcommand{\Vkon}{\ensuremath{ \,V_{_{k+1}}^n }}
\newcommand{\Vki}{\ensuremath{{ \,V_{_{k+i}} }}}

\newcommand{\Vz}{\ensuremath{{ \,V^0 }}}
\newcommand{\Vzko}{\ensuremath{{ \,\Vz_{_{\hspace{-1.5ex}k+1}} }}}

\newcommand{\Vbar}{\ensuremath{ \overline{V} }}
\newcommand{\Vbark}{\ensuremath{ \Vbar_{_{k}} }}

\newcommand{\Wk}{\ensuremath{{ \,W_{_k} }}}
\newcommand{\Wz}{\ensuremath{{ \,W_{_0} }}}
\newcommand{\Wzz}{\ensuremath{ \,W^{0}_{_0} }}
\newcommand{\Wzo}{\ensuremath{ \,W^{1}_{_0} }}
\newcommand{\Wo}{\ensuremath{{ \,W_{_1} }}}
\newcommand{\Wko}{\ensuremath{{ \,W_{_{k+1}} }}}
\newcommand{\Wki}{\ensuremath{{ \,W_{_{k+i}} }}}
\newcommand{\Woj}{\ensuremath{{ \,W_{_{1j}} }}}
\newcommand{\Woi}{\ensuremath{{ \,W_{_{1i}} }}}
\newcommand{\Wnk}{\ensuremath{{ \,Wn_{_k} }}}
\newcommand{\Wi}{\ensuremath{{ \,W_{_i} }}}
\newcommand{\WN}{\ensuremath{{ \,W_{_N} }}}
\newcommand{\WNo}{\ensuremath{ \,W_{_{N+1}} }}
\newcommand{\Wkt}{\ensuremath{ W_{_k}^o }}

\newcommand{\Xk}{\ensuremath{{ \,X_{_k} }}}
\newcommand{\Xko}{\ensuremath{{ \,X_{_{k+1}} }}}
\newcommand{\Xz}{\ensuremath{{ \,X_{_0} }}}
\newcommand{\Xinf}{\ensuremath{{ \,X_{_\infty} }}}

\newcommand{\Xkk}{\ensuremath{{ \,{\widehat{X}}_{_{k/k}} }}}
\newcommand{\Xkok}{\ensuremath{{ \,{\widehat{X}}_{_{k+1/k}} }}}
\newcommand{\Xkoko}{\ensuremath{{ \,{\widehat{X}}_{_{k+1/k}} }}}
\newcommand{\Xzz}{\ensuremath{{ \,{\widehat{X}}_{_{0/0}} }}}
\newcommand{\Xer}{\ensuremath{{ \,{\widetilde{X}} }}}

\newcommand{\Xt}{\ensuremath{{ \,X_{_t} }}}

\newcommand{\Xik}{\ensuremath{ \Xi_{_k} }}
\newcommand{\Xiko}{\ensuremath{ \Xi_{_{k+1}} }}
\newcommand{\Xiz}{\ensuremath{{ \Xi(\qz) }}}
\newcommand{\Xio}{\ensuremath{{ \Xi(\qo) }}}
\newcommand{\Xiezz}{\ensuremath{{ \Xi(\qezz) }}}
\newcommand{\Xieoo}{\ensuremath{{ \Xi(\qeoo) }}}
\newcommand{\Xie}{\ensuremath{ \,\widehat{\Xi} }}
\newcommand{\Xiek}{\ensuremath{ \Xie_{_k} }}
\newcommand{\Xieko}{\ensuremath{ \Xie_{_{k+1}} }}
\newcommand{\Xiekok}{\ensuremath{ \Xie_{_{k+1/k}} }}
\newcommand{\Xiektwk}{\ensuremath{ \Xie_{_{k+2/k}} }}
\newcommand{\Xiekoko}{\ensuremath{ \Xie_{_{k+1/k+1}} }}
\newcommand{\Xiekolk}{\ensuremath{ \Xie_{_{k+l-1/k}} }}
\newcommand{\Xieklk}{\ensuremath{ \Xie_{_{k+l/k}} }}
\newcommand{\Xiekk}{\ensuremath{ \Xie_{_{k/k}} }}

\newcommand{\Yt}{\ensuremath{{ Y_{_t} }}}

\newcommand{\Yko}{\ensuremath{{ \,Y_{_{k+1}} }}}

\newcommand{\zvcross}{\ensuremath{ \,\left[ \zv \times \right] }}
\newcommand{\zbcross}{\ensuremath{ \,\left[ \zb \times \right] }}
\newcommand{\zbicross}{\ensuremath{ \,\left[ \zbi \times \right] }}
\newcommand{\zbncross}{\ensuremath{ \,\left[ \zbn \times \right] }}
\newcommand{\zkocross}{\ensuremath{ \,\left[ \zko \times \right] }}
\newcommand{\zepscross}{\ensuremath{ \,\left[ \zeps \times \right] }}

\newcommand{\Zt}{\ensuremath{{ Z_{_t} }}}

\newcommand{\Zthree}{\ensuremath{{ 0_{3\times3} }}}


\newcommand{\AD}{\ensuremath{{ \bf{AD}\,  }}}
\newcommand{\EMF}{\ensuremath{{ \bf{EMF}\,  }}}
\newcommand{\EKF}{\ensuremath{{ \bf{EKF}\,  }}}
\newcommand{\RLS}{\ensuremath{{ \bf{RLS}\,  }}}
\newcommand{\KVKF}{\ensuremath{{ \bf{KVKF}\,  }}}
\newcommand{\quatKF}{\ensuremath{{ \bf{quatKF}\,  }}}
\newcommand{\REQUEST}{ \mbox{REQUEST} }

\newcommand{\xnorm}{\ensuremath{ \, \| \xv \|  }}
\newcommand{\xonorm}{\ensuremath{ \, \| \xo \|  }}
\newcommand{\xtwnorm}{\ensuremath{ \, \| \xtw \|  }}
\newcommand{\xounit}{\ensuremath{ \,\frac{\xo}{\| \xo \|}  }}
\newcommand{\axsum}{\ensuremath{  \,\left( \ao\,\xo\;+\;\atw\,\xtw \right)  }}
\newcommand{\axsumnorm}{\ensuremath{  \|\ao\,\xo\;+\;\atw\,\xtw \| }}
\newcommand{\aoatwsum}{\ensuremath{ \,\left( \ao + \atw \right)  }}

\newcommand{\ipo}{\ensuremath{ \,\mbox{ip}_{_1} }}
\newcommand{\iptw}{\ensuremath{ \,\mbox{ip}_{_2} }}
\newcommand{\ipmoo}{\ensuremath{ \,\mbox{ipm}_{_{11}} }}
\newcommand{\ipmtt}{\ensuremath{ \,\mbox{ipm}_{_{22}} }}

\newcommand{\pcmoo}{\ensuremath{ \,\mbox{pcm}_{_{11}} }}
\newcommand{\pcmtt}{\ensuremath{ \,\mbox{pcm}_{_{22}} }}
\newcommand{\gainoo}{\ensuremath{ \,\mbox{gain}_{_{11}} }}
\newcommand{\gaintt}{\ensuremath{ \,\mbox{gain}_{_{22}} }}

\begin{authorList}{4cm}
 \addAuthor{Caitong Peng}{Graduate Student, Ben-Gurion University of the Negev, Mechanical Engineering dept., 84105, Beer Sheva, Israel. \emailAddress{caitongpeng@gmail.com}}

\addAuthor{Daniel Choukroun}{Senior Lecturer, Ben-Gurion University of the Negev, Mechanical Engineering dept., 84105, Beer Sheva, Israel. \emailAddress{danielch@bgu.ac.il}}

\end{authorList}


\begin{abstract}
A novel single-frame quaternion estimator processing two vector observations is introduced. The singular cases are examined, and appropriate rotational solutions are provided. Additionally, an alternative method involving sequential rotation is introduced to manage these singularities. The simplicity of the estimator enables clear physical insights and a closed-form expression for the bias as a function of the quaternion error covariance matrix. The covariance could be approximated up to second order with respect to the underlying measurement noise assuming arbitrary probability distribution. The current note relaxes the second-order assumption and provides an expression for the error covariance that is exact to the fourth order, under the assumption of Gaussian distribution. A comprehensive derivation of the individual components of the quaternion additive error covariance matrix is presented. This not only provides increased accuracy but also alleviates issues related to singularity.
 
\end{abstract}

\keywords{Attitude quaternion, sequential rotations, Bias, Covariance, multivariate Gaussian distribution}

\section{Introduction}\label{sec1}

Attitude determination is critical to many aerospace missions and has several decades of history. The quaternion of rotation~\cite[p. 758]{wrz}, a singularity-free minimal attitude representation, is known to present excellent numerical and analytical properties and has become very popular for designing attitude estimators. Quaternion attitude estimators may be classified as single-frame algorithms or filters. The latter class typically stems from optimal stochastic filtering theory and allows tracking of time-varying attitude parameters as well as other states like sensor errors. It operates via incremental changes and is highly sensitive to a priori statistical knowledge of the noises. Single frame estimators, on the other hand, born in the realm of constrained deterministic least-squares theory, lend themselves to batch algorithms, and were extended to recursive versions, including time-varying attitude and additional parameters. An appealing advantage of single-frame attitude estimators over filters is the lack of sensitivity to initial conditions since they provide global rather than incremental estimates. Rooted in the Wahba problem~\cite{wba}, the quaternion batch algorithm known as the q-method, early reported in \cite{kit}, has given rise to sophisticated versions aiming at providing closed-form solutions and reducing the computational burden, e.g. \cite{shu1, mor1, mor2} to cite a few. An error analysis of the q-method is revisited and extended to non-unit and noisy reference vectors in \cite{cha}. Excellent surveys of algorithms and error analyses can be found in \cite{cra1} and \cite[Chap. 5]{mrk1}. The q-method and related algorithms involve the solution of symmetric eigenvalues problems in dimension four, with computation burdens that naturally increase with the number of observations. While it seems efficient to process as many measurements as available, the insight in solutions for more than two observations is lost, and the marginal increase in accuracy might not pay off. Furthermore, there is a growing class of very small satellites, a.k.a. nanosatellites or CubeSats, that accommodate two sensors only onboard, e.g. a magnetometer and a Sun sensor. Early works in this realm are still used nowadays. The TRIAD estimator uses exactly two vector measurements and devises a virtual third one via an orthogonalization process in order to estimate the attitude matrix~\cite{blk}. A covariance analysis of TRIAD for a multiplicative error in terms of the Euler vector is presented in~\cite{shu1}. A generalized TRIAD algorithm is introduced in ~\cite{bar1} via gain optimization. Ref.~\cite{mrk2} presents an optimal attitude matrix estimator from two vector measurements in the realm of the Wahba problem. In \cite{shu1} a closed-form expression of the QUEST algorithm is developed for the case of two observations. In~\cite{mor3} the EULER 2 estimator utilizes the Rodrigues formula for reconstructing the Euler axis/angle parameters and the attitude matrix. It invokes optimization of Wahba's loss function in the case of noisy measurements. It exploits a coplanarity condition of the measured and predicted vectors towards the development of analytical formulas. Ref.~\cite{rey1} introduces a quaternion parameterization given a single vector observation. The degree of freedom is an angle around the observation. Several methods are devised for analytical determination of the attitude quaternion given two observations. Interestingly, the quaternions were manipulated as classes of equivalence where the elements are collinear but not necessarily unit-norm. Motivated by the insight provided in~\cite{rey1}, an optimization method on the degrees of freedom in the quaternion parametrizations related to two observations is devised in \cite{mrk3}. Very efficient algorithms emerged: an optimal one, more accurate and as fast as TRIAD, and a suboptimal one, faster and as accurate as TRIAD. 

The QUEST algorithm~\cite{shu1} requires solving for the maximum eigenvalue in a characteristic equation, which is known to be less robust than more direct approaches in computing the associated eigenvector. The related computation burden is greater than other similar quaternion estimators. The estimator in~\cite{mor3} isn't properly a quaternion estimator (neither are the TRIAD and related attitude matrix estimators), it involves trigonometry, which burdens the computation load, and lacks singular cases mitigation. In~\cite{rey1}, the first method seeks a quaternion in the span of two base quaternions but no estimator is actually devised. While the second method exploits geometrical insights for developing a quaternion estimate, it breaks down in specific unaddressed singular cases. The work in~\cite{mrk3} addresses the singularities, and its analytical formulas for the quaternion help with the computation load, but the optimization still adds to the calculations. More strikingly, none of the above-mentioned works feature some error analysis, in particular for the bias and covariance of the estimation error. A covariance analysis of the q-method is presented in ~[Chap. 5]\cite{mrk1}. Under the typical assumptions of unbiasedness and first-order approximations in the measurement noises, the estimation errors are unbiased. This raises the question of whether biases and covariance could be analytically expressed, for additive or multiplicative quaternion errors, and of whether better estimation accuracy could be obtained as a result. 

This work is concerned with the development and analysis of a very fast quaternion estimator from two vector observations. It thoroughly addresses the singularity cases and addresses them via sequential rotations. A deterministic error analysis is performed that lends itself to analytical expressions for the biases and covariances of the quaternion multiplicative and additive errors. Using results from~\cite{chk3, chk2}, the quaternion is sought as the unique solution to a set of orthogonality conditions that involve simple nonlinear expressions of the measurements. This approach fundamentally differs from \cite{rey1, mrk3} because the quaternion is not sought in the span of a particular basis, and from \cite{mor1} because the orthogonality conditions do not require solving for the optimal Wahba's loss value. The proposed algorithm is exceedingly simple, which dramatically simplifies the computations. This also enables a systematic analysis and mitigation of singularity cases. It is carried out using sequential rotations, where, as noted in \cite{mrk3}, it is preferable to choose a single desirable rotation as early as possible in order to save computations. The simplicity of the estimator also enables a thorough deterministic and random error analysis. It sheds additional light on the various nonlinear effects in quaternion estimation, including normalization. The analysis is carried out in four dimensions both for multiplicative and additive errors and lends itself to second-order and fourth-order expressions for the biases and for the errors covariance matrices, respectively, in terms of the measurement noises. 

The remainder of the paper is organized as follows. Section~\ref{sec2} presents quaternion estimation using two vector observations.
Section~\ref{sec3} addresses the singularity cases. Section~\ref{sec4} is concerned with the error analysis.
Section~\ref{sec5} expressions accurate to the fourth order are developed in the case of Gaussian noise.
And the conclusions are drawn in Section~\ref{sec7}.


\section{Quaternion Estimation using Two Vector Observations}\label{sec2}

\subsection{Preliminaries}
This section follows Ref.~\cite{chk2}. Let $\bv$ and $\rv$ denote the projections of an ideal noise-free
vector measurement on a body coordinate frame and a reference coordinate frame, $\Bfr$ and $\Rfr$, respectively. The rotation quaternion from $\Rfr$ to $\Bfr$, denoted by $\qv$, belongs to the null space of the following matrix:
\begin{align}
\label{s2eq00}
 & H =
 \begin{bmatrix}
 - \scross & \dv \\
 -\dv^T & 0 \\
 \end{bmatrix} \
\end{align}
where
\begin{align}
\label{s2eq01}
& \sv = \half (\bv+\rv) \\
\label{s2eq02} 
& \dv = \half (\bv-\rv) \
\end{align}
and $\scross$ denotes the cross-product matrix built from the $3 \times 1$ vector $\sv$. The spectral decomposition of the matrix $H$ features a kernel, $\kernel{H}$, that is generated by the orthonormal basis $\{\qonev, \qtwov\}$ where
\begin{align}
\label{s2eq04}
 & \qonev =
 \begin{bmatrix}
 \sv \\
 0 \\
 \end{bmatrix} \frac{1}{\|\sv\|} \\
\label{s2eq05}
 & \qtwov =
 \begin{bmatrix}
 - \sv \times \dv \\
 \|\sv\|^2 \\
 \end{bmatrix}\,
 \frac{1}{\|\sv\|} \
\end{align}

In addition, the orthogonal complement plane to $\kernel{H}$, $(\kernel H)^{\bot}$, is generated by the orthonormal basis $\{\qthrv, \qforv\}$ where
\begin{align}
\label{s2eq04a}
 & \qthrv =
 \begin{bmatrix}
 \dv \\
 0 \\
 \end{bmatrix} \frac{1}{\|\dv\|} \\
\label{s2eq05a}
 & \qforv =
 \begin{bmatrix}
 \sv \times \dv \\
 \|\dv\|^2 \\
 \end{bmatrix}\,
 \frac{1}{\|\dv\|} \
\end{align}

Both $\qonev$ and $\qtwov$ are feasible candidates to represent the rotation from $\Rfr$ to $\Bfr$ since they both belong to the null space of $H$. The quaternion $\qonev$ is characterized by a rotation angle of 180 degrees while $\qtwov$ features a minimum angle. One may seek the true quaternion as a linear combination of $\qonev$ and $\qtwov$. In this work, however, we follow an orthogonal route.

\subsection{Closed-form quaternion estimator}
Given two ideal non-collinear vector observations associated with the same attitude, i.e., two pairs of noise-free unit-norm column-vectors, $(\bonev,\ronev)$ and $(\btwov,\rtwov)$, such that the angle between $\ronev$ and $\rtwov$ equals the angle between $\bonev$ and $\btwov$, then the true quaternion $\qv$ is expressed as follows:
\begin{equation}
\label{s3eq00}
 \qv =
 \begin{bmatrix}
 \donev \times \dtwov \\
 \sonev^T \dtwov \\
 \end{bmatrix}
 \,\frac{1}{\sqrt{\|\donev\times\dtwov\|^2 + |\sonev^T\dtwov|^2}}
\end{equation}
where
\begin{align}
\label{s3eq00a}
& \siv = \half (\biv + \riv), \;\;\; i=1,2
 \\ 
 \label{s3eq00b}
& \dv_{_{i}} = \half (\biv - \riv), \;\;\;  i=1,2 \
\end{align}

The proof follows. Let $\qijv$ denote the $j^{th}$ basis element, $j=1,2,3,4$, constructed from the $i^{th}$ vector measurement, $i=1,2$. As noted in Section~\ref{sec2}, the quaternion $\qv$ belongs to the orthogonal complements of the two planes generated by the pairs $(\qonethrv,\qoneforv)$ and $(\qtwothrv,\qtwoforv)$, respectively.  The sought quaternion can thus be found along the intersection of these planes. Let $\xv$ denote a feasible unnormalized quaternion with vector part $\av$ and scalar part $\alpha$, then the orthogonality relationships yield
\begin{align}
\label{s3eq01}
& \donev^T \av = 0 \\
\label{s3eq02}
& \dtwov^T \av = 0 \\
\label{s3eq03}
& (\sonev \times \donev)^T \av + \|\donev\|^2 \alpha = 0 \\
\label{s3eq04}
 & (\stwov \times \dtwov)^T \av + \|\dtwov\|^2 \alpha = 0 \
\end{align}

Eqs.~\eqref{s3eq01}-\eqref{s3eq04} are linearly dependent otherwise $\qv$ would be null. For simplicity, we will use Eqs.~\eqref{s3eq01}-\eqref{s3eq03}, only. Eqs.~\eqref{s3eq01}-\eqref{s3eq02} clearly show that a feasible choice for $\av$ is:
\begin{equation}
\label{s3eq05}
 \av = \donev \times \dtwov
\end{equation}
so that, using Eq.~\eqref{s3eq05} in Eq.~\eqref{s3eq03},  $\alpha$
is determined as follows:
\begin{align*}
 \nonumber
 (\sonev \times \donev)^T (\donev \times \dtwov) + \|\donev\|^2
 \alpha & = 0 \\
 \nonumber
 \sonev^T \donevcross \donevcross \dtwov + \|\donev\|^2 \alpha & = 0
 \\
 \nonumber
 (\sonev^T \donev) \donev^T\dtwov - \|\donev\|^2 \sonev^T \dtwov + \|\donev\|^2 \alpha & =
 0 \\ 
 \nonumber
 \|\donev\|^2 (\alpha - \sonev^T \dtwov) & = 0 \\
 \label{s3eq06}
  \alpha & = \sonev^T \dtwov \
\end{align*}
which, upon normalization of the quaternion, yields the sought
result. Finally, notice that due to the easily verifiable identity:
\begin{equation*}
 \sonev^T \dtwov = -\stwov^T \donev
\end{equation*}
an identical expression for $\qv$ is readily obtained by using
Eq.~\eqref{s3eq04} instead of Eq.\eqref{s3eq03}.

 \section{Singular Cases} \label{sec3}

The proposed estimator is exceedingly simple but it sometimes yields singular expressions, i.e. the quaternion can not be fully determined. This section reviews the singular cases, sheds light on the geometry configurations, and provides the actual solution. A straightforward inspection of the estimator's Eqs. \eqref{s3eq00} yields the following singularity cases, as summarized in Tab.~\ref{tab1}.
\begin{table}[H]
 \centering
 \caption{Singular cases and the associated rotations}
 \label{tab1}
 \vspace{4mm}
\begin{tabular}{@{\extracolsep{10mm}}ccl}
\hline\hline
Case    &   Source   &   Rotation type \\

\hline

A & $\donev = \dtwov = 0$ & Zero attitude \\ 
B & $\donev = 0$ and $\dtwov \neq 0$ & Rotation around first VM, Eq.~\eqref{s4eq06} \\
C & $\donev \neq 0$ and $\dtwov = 0$ & Rotation around second VM, Eq.~\eqref{s4eq06a} \\
D & $\donev \times \dtwov = 0$ and $\donev \neq 0, \dtwov \neq 0$ & Rotation around $\uv$ with angle $\alpha$, Eqs.~\eqref{s4eq09},\eqref{s4eq10}\\
\hline\hline
\end{tabular}
\end{table}
\subsection*{Case A}

Case A is defined when both vectors $\donev$ and $\dtwov$ are zero, which means that the body and reference vectors are identical, i.e.
\begin{align}
\label{sec4eq00}
    & \bonev = \ronev \\
\label{sec4eq01}
    & \btwov = \rtwov \
\end{align}

This is satisfied if and only if the frames $\Bfr$ and $\Rfr$ coincide, i.e. when the attitude is identity.  

\subsection*{Case B}

\begin{figure}[H]
\begin{center} \resizebox{5cm}{5cm}{
\includegraphics{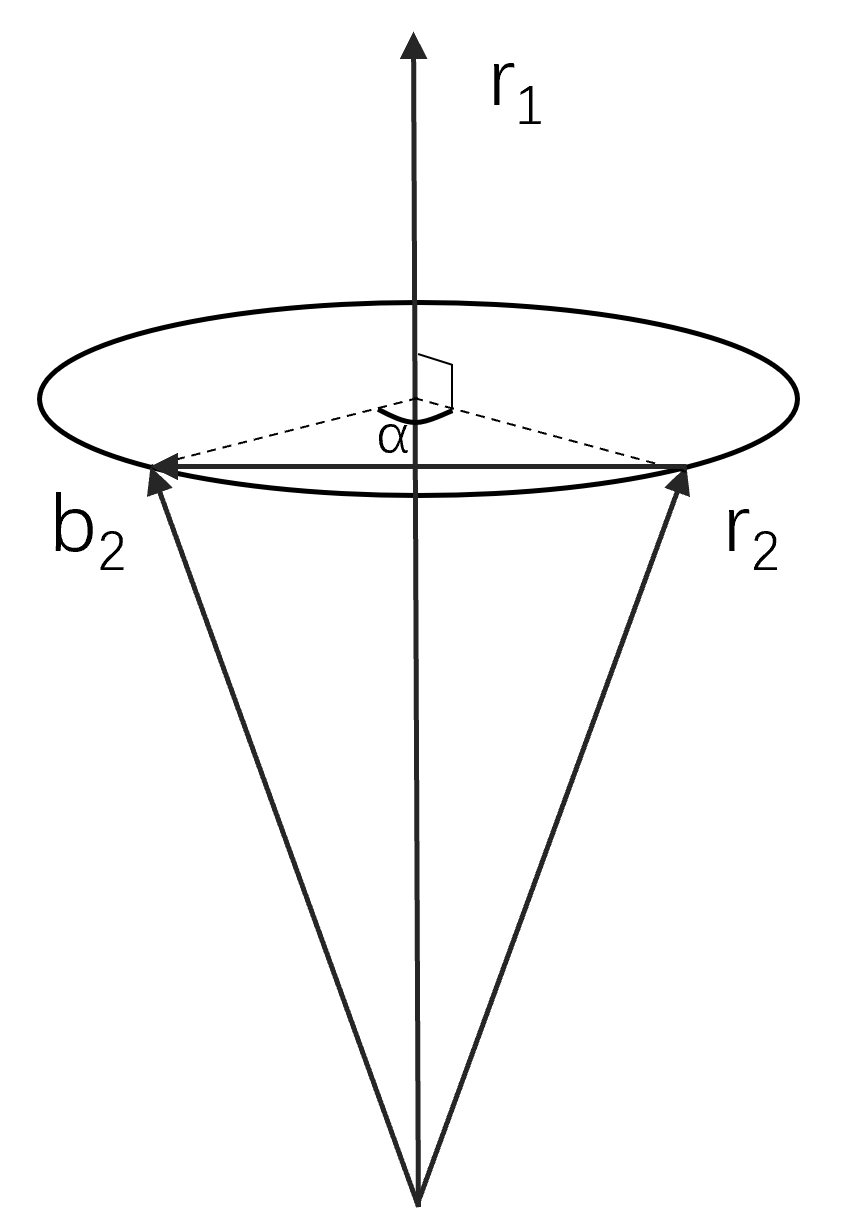}}
\caption{Case B. Rotation around VM 1.} \label{s4f1}
\end{center}
\end{figure}

\begin{figure}[H]
\begin{center} \resizebox{10cm}{4cm}{
\includegraphics{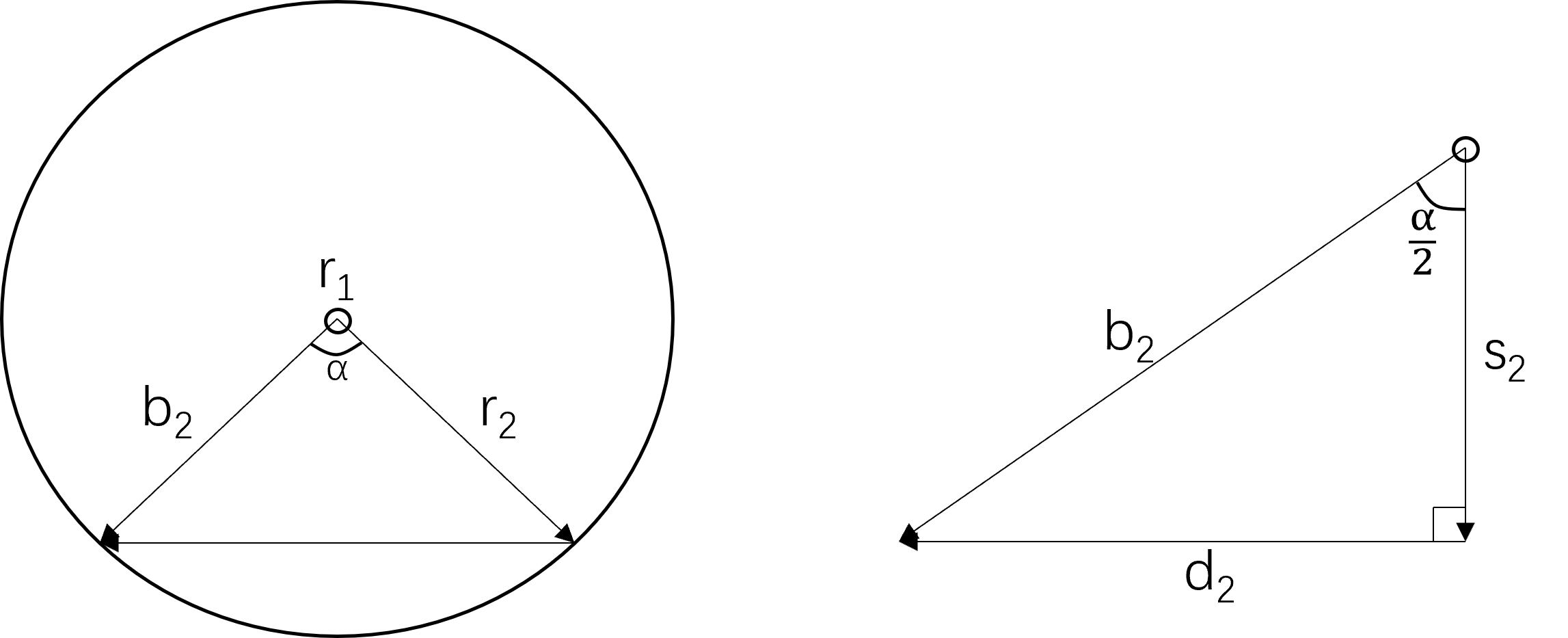}}
\caption{Case B. Rotation around VM 1. View from above.} \label{s4f2}
\end{center}
\end{figure}

This case stems from the cancellation of $\donev$ only. We deduce from Eq.~\eqref{sec4eq00} that the first vector measurement (VM 1) is invariant. This provides the unit vector of the rotation axis, here $\ronev$. Notice that the difference vector $\dtwov$ is necessarily perpendicular to the rotation axis, as illustrated in Fig.~\ref{s4f1}. 
This is true for an arbitrary rotation angle $\alpha$. The latter can be determined unless both measurements are collinear. Assume for simplicity that they are perpendicular. The quaternion of rotation that brings $\btwov$ to $\rtwov$ via a rotation around $\ronev$ is the minimum angle rotation, as shown by inspecting Fig.~\ref{s4f2}. The drawing on the left-hand side depicts the rotation from above and the drawing on the right-hand side focuses on the right-angle triangle whose sides are the vectors $\stwov$ and $\dtwov$, with hypotenuse of length $\|\btwov\|$, i.e. equals 1.    
Inspection of the right-angle triangle yields the following identities:
\begin{align}
\label{s4eq04}
& \sin{\frac{\alpha}{2}} = \|\dtwov\| \\
\label{s4eq05}
& \cos{\frac{\alpha}{2}} = \|\stwov\| \
\end{align}

Recalling that the quaternion of rotation is defined as follows:
\begin{align}
\label{s4eq02}
& \qv = 
\begin{bmatrix}
    \ronev\;\sin{\frac{\alpha}{2}} \\ \cos{\frac{\alpha}{2}} \\
\end{bmatrix} \
\end{align}
and substituting Eqs.~\eqref{s4eq04}- \eqref{s4eq05} into Eq.~\eqref{s4eq02} yields the following quaternion:
\begin{align}
\label{s4eq06}
& \qv = 
\begin{bmatrix}
    \ronev \|\dtwov\| \\ \|\stwov\| \\
\end{bmatrix} \
\end{align}

If the VMs are not perpendicular, then we first create a pseudo-measurement as the cross-product of the two VMs, substitute it to the second VM, and proceed identically to the above. In addition, it is straightforward to show that the following identity holds:
\begin{align}
\label{s4eq07}
 \ronev \|\dtwov\| = \frac{\dtwov \times \stwov}{\|\stwov\|} \
\end{align}
where the right-hand side is identical to the vector part of the minimal-angle quaternion built from a single VM (see Eq.~\eqref{s2eq05}). To conclude, this singular case corresponds to a rotation around the first vector measurement with a minimum angle. In the particular case of a 180-degree rotation, the Eq. \eqref{s2eq05} breaks down while Eq.~\eqref{s4eq06} remains valid.

\subsection*{Case C}

Case C is defined when the difference vector $\dtwov$ cancels out. 
By symmetry with case B one concludes that the rotation is around the second vector measurement. If the measurements are perpendicular, the quaternion is determined as 
\begin{align}
\label{s4eq06a}
& \qv = 
\begin{bmatrix}
    \rtwov \|\donev\| \\ \|\sonev\| \\
\end{bmatrix} \
\end{align}

Otherwise, the first VM is replaced by the normalized cross-product of the two VMs.

\subsection*{Case D}

\begin{figure}[H]
\begin{center} \resizebox{11cm}{5cm}{
\includegraphics{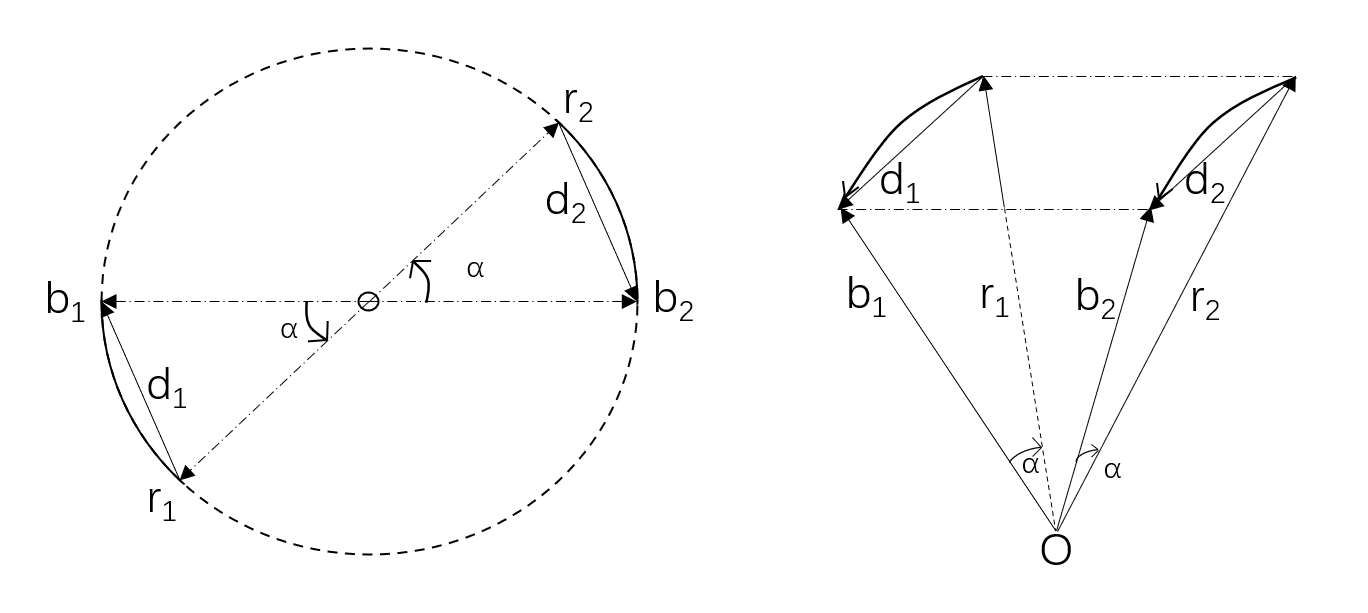}}
\caption{Case D. Rotation that maintains $\donev$ and $\dtwov$ parallel.} \label{s4f3}
\end{center}
\end{figure}

\begin{figure}[H]
\begin{center} \resizebox{5.5cm}{3.5cm}{
\includegraphics{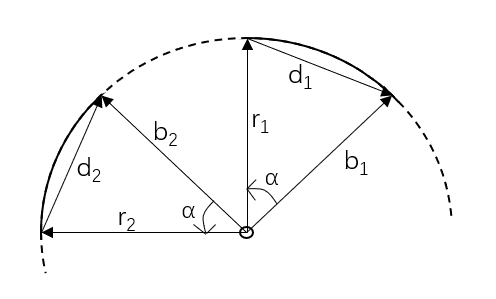}}
\caption{Rotation where $\donev$ and $\dtwov$ are not parallel.} \label{s4f4}
\end{center}
\end{figure}

This case is characterized by the identity
\begin{align}
\label{s4eq08}
& \donev \times \dtwov = 0 \
\end{align}
that happens when the vector differences $\donev$ and $\dtwov$ are parallel. Consider the plan spanned by the pair $(\bonev,\btwov)$. A general rotation of this plane around a fixed point $O$ can be decomposed into two axial rotations: the first one around $(\bonev \times \btwov)$ and the other around an axis lying in the plane $(\bonev,\btwov)$. An interesting feature of the latter rotation is that it creates vector differences $(\donev,\dtwov)$ that remain parallel to each other, see Fig.~\ref{s4f3}. The former rotation on the other hand necessarily breaks the parallelism between $\donev$ and $\dtwov$, see Fig.~\ref{s4f4}. The left-hand side in Fig.~\ref{s4f3} illustrates the case of a rotation axis along $(\bonev+\btwov)$: $\donev$ and $\dtwov$ are equal and opposite in direction. The right-hand side pictures the case of a rotation along $(\bonev - \btwov)$: $\donev$ and $\dtwov$ are identical. The general case is a composition of these two.  

Bringing the triangle $(O,\bonev,\btwov)$ onto the triangle $(O,\ronev,\rtwov)$ may be accomplished by a rotation around the line intersection of both planes. The rotation quaternion can be uniquely expressed as a function of the pair $(\uv, \alpha)$, the eigenaxis and angle, respectively, that are expressed as follows:
\begin{align}
\label{s4eq09}
& \uv = \frac{\bthrv \times \rthrv}{\| \bthrv \times \rthrv\|} \\
\label{s4eq10}
& \cos \alpha = \bthrv^T\rthrv \
\end{align}
where $\alpha \in [0^\circ, 180^\circ]$ and
\begin{align}
\label{s4eq09a}
& \bthrv = \frac{\bonev \times \btwov}{\| \bonev \times \btwov\|}\\
\label{s4eq10a}
& \rthrv = \frac{\ronev \times \rtwov}{\| \ronev \times \rtwov \|} \
\end{align}

To conclude, this singular case corresponds to a rotation that can be unambiguously determined from the single pseudo-vector measurement formed by the cross-product of the two original measurements. There is no contradiction with the fact that two vector measurements are required for three-dimensional attitude determination. In the present case, the ambiguity due to an unknown rotation around the single VM does not exist owing to the parallelism constraint.

\subsection{Sequential rotations}

In practice singularity seldom happens because the noise in the data often prevents perfect cancellations of the vector differences or their parallelism. Yet, when the noises are very low, the estimator's formula may be badly conditioned. This difficulty can be circumvented by the method of sequential rotations~\cite[p. 192]{mrk1}. The general approach is explained as follows and the particular sequences are proposed next. Assuming that the true attitude from $\Rfr$ to $\Bfr$ yields one of the singular cases, then a new reference frame $\Cfr$ is sought such that the rotation from $\Cfr$ to $\Bfr$ yields well-behaved estimation ${\qev^{\Cfr}}_{\Bfr}$. Once the quaternion ${\qev^{\Cfr}}_{\Bfr}$ is estimated, the quaternion from $\Rfr$ to $\Bfr$ may be retrieved by a simple composition with the quaternion from $\Rfr$ to $\Cfr$. Typically, the transformation  from $\Rfr$ to $\Cfr$ and rotations of $\pi$ radians about one of the standard axes are as follows: ${\qev^{\Rfr}}_{\Bfr_{R(\xv, \pi)}} =[q_4,-q_3,q_2,-q_1]^T$,
${\qev^{\Rfr}}_{\Bfr_{R(\yv, \pi)}} =[q_3,q_4,-q_1,-q_2]^T$,
${\qev^{\Rfr}}_{\Bfr_{R(\zv, \pi)}} =[-q_2,q_1,q_4,-q_3]^T$, where $R(\xv, \pi)$ means rotate $\pi$ along with the $\xv$ axis and $q_4$ and $q_{1:3}$ are scalar and vector part of the quaternion. Tab.~\ref{tab2} summarizes these validation conditions.

\begin{table}[H]
 \centering
 \caption{Validation criteria for simple rotations}
 \label{tab2}
 \scalebox{0.9}{
\begin{tabular}{@{\extracolsep{10mm}}ll}
\hline\hline
Singular Case    &  Validation Criteria
\\
\hline
&$R(\xv, \pi)$ invalid IF  $\bonev$ $\parallel$ x-axis OR $\btwov$ $\parallel$ x-axis OR 
$[ 0 ,r_{1y} ,r_{1z}  ] \parallel [0 ,r_{2y} ,r_{2z}  ]$ \\
A.$\donev = \dtwov = 0$&$R(\yv, \pi)$ invalid IF   $\bonev$ $\parallel$ y-axis OR $\btwov$ $\parallel$ y-axis OR $[
r_{1x} ,0 ,r_{1z}  ] \parallel [r_{2x} ,0 ,r_{2z}  ] $ \\
     &$R(\zv, \pi)$ invalid IF   $\bonev$ $\parallel$ z-axis OR $\btwov$ $\parallel$ z-axis OR $[
r_{1x} ,r_{1y} ,0  ] \parallel [r_{2x} ,r_{2y} ,0  ] $ \\
 \hline 
    &$R(\xv, \pi)$ invalid IF  $\bonev$ $\parallel$ x-axis AND $(b_{2y}+r_{2y})r_{1z}=k(b_{2z}+r_{2z})r_{1y}$\\
B.$\donev = 0$ \& $\dtwov \neq 0$ &$R(\yv, \pi)$ invalid IF  $\bonev$  $\parallel$ y-axis AND $(b_{2x}+r_{2x})r_{1z}=k(b_{2z}+r_{2z})r_{1x}$\\
    &$R(\zv, \pi)$ invalid IF  $\bonev$  $\parallel$ z-axis AND $(b_{2y}+r_{2y})r_{1x}=k(b_{2x}+r_{2x})r_{1y}$\\
\hline
    &$R(\xv, \pi)$ invalid IF  $\btwov$ $\parallel$ x-axis AND $(b_{1y}+r_{1y})r_{2z}=k(b_{1z}+r_{1z})r_{2y}$ \\
C.$\donev \neq 0$ \& $\dtwov = 0$ &$R(\yv, \pi)$ invalid IF  $\btwov$ $\parallel$ y-axis AND $(b_{1x}+r_{1x})r_{2z}=k(b_{1z}+r_{1z})r_{2x}$\\
    &$R(\zv, \pi)$ invalid IF  $\btwov$ $\parallel$ z-axis AND $(b_{1y}+r_{1y})r_{2x}=k(b_{1x}+r_{1x})r_{2y}$\\
\hline
    &$R(\xv, \pi)$ invalid IF 
    $\donev + [ 0 ,r_{1y} ,r_{1z}  ]  
    \parallel 
    \dtwov + [0 ,r_{2y} ,r_{2z}  ]$\\
D.$\donev \times \dtwov = 0$ \& $\donev \neq 0, \dtwov \neq 0$ &$R(\yv, \pi)$ invalid IF 
$\donev + [ r_{1x} ,0 ,r_{1z}  ] 
\parallel
\dtwov + [r_{2x} ,0 ,r_{2z}  ]$\\
    &$R(\zv, \pi)$ invalid IF 
    $\donev +[ r_{1x} ,r_{1y} ,0  ]
    \parallel
    \dtwov + [r_{2x} ,r_{2y} ,0   ]$\\

\hline\hline
\end{tabular}}
\end{table}

\textbf{The development of simple validation criteria for each permutation is presented as follows:}
 
\textbf{Case A: $\donev = \dtwov = 0$}
Assume a rotation of $\pi$ radians about the x-axis:
\begin{align*}
    &\bonev=\ronev=[ r_{1x},r_{1y},r_{1z}]^T\\
    &\btwov=\rtwov=[ r_{2x},r_{2y},r_{2z}]^T\\
    &\ronev^{\Cfr}=[r_{1x},-r_{1y},-r_{1z}]^T\\
    &\rtwov^{\Cfr}=[r_{2x},-r_{2y},-r_{2z}]^T\\
    &\donev^{\Cfr}=[0 ,r_{1y} ,r_{1z} ]^T\\
    &\dtwov^{\Cfr}=[0 ,r_{2y} ,r_{2z} ]^T
\end{align*}

If $\bonev$ $\parallel$ x-axis or $\btwov$ $\parallel$ x-axis or $[0,r_{1y},r_{1z}]^T \parallel [0,r_{2y},r_{2z} ]^T$, then the vector differences $\donev^{C}, \dtwov^C$ are in one of the singular cases. Similar conclusions are readily obtained for rotations around the other two axes by $\pi$ radians. 

\textbf{Case B: $\donev = 0$ \& $\dtwov \neq 0$}
Assume a rotation of $\pi$ radians about the x-axis:
\begin{align*}
    &\bonev=\ronev=[r_{1x},r_{1y},r_{1z}]^T\\
    &\rtwov=[r_{2x},r_{2y},r_{2z}]^T\\
    &\btwov=[b_{2x},b_{2y},b_{2z}]^T\\
    &\ronev^{\Cfr}=[r_{1x},-r_{1y},-r_{1z}]^T\\
    &\rtwov^{\Cfr}=[r_{2x},-r_{2y},-r_{2z}]^T\\
    &\donev^{\Cfr}=[0,r_{1y},r_{1z} ]^T\\
    &\dtwov^{\Cfr}=\half [ b_{2x}-r_{2x},b_{2y}+r_{2y},b_{2z}+r_{2z}]^T
\end{align*}

If $\bonev$ $\parallel$ x-axis and $(b_{2y}+r_{2y})r_{1z}=k(b_{2z}+r_{2z})r_{1y}$, where $k$ is an arbitrary constant, then the vector differences $\donev^{C}$ and $\dtwov^C$ are in one of the singular cases. Similar conclusions are readily obtained for rotations around the other two axes by $\pi$ radians.

\textbf{Case C: $\donev \neq 0$ \& $\dtwov = 0$}
Assume a rotation of $\pi$ radians about the x-axis:
\begin{align*}
    &\ronev=[r_{1x},r_{1y},r_{1z}]^T\\
    &\bonev=[b_{1x},b_{1y},b_{1z}]^T\\
    &\btwov=\rtwov=[r_{2x},r_{2y},r_{2z}]^T\\
    &\ronev^{\Cfr}=[r_{1x},-r_{1y},-r_{1z}]^T\\
    &\rtwov^{\Cfr}=[r_{2x},-r_{2y},-r_{2z}]^T\\
    &\donev^{\Cfr}=\half [ b_{1x}-r_{1x},b_{1y}+r_{1y},b_{1z}+r_{1z}]^T\\
    &\dtwov^{\Cfr}=[0,r_{2y},r_{2z} ]^T
\end{align*}

If $\btwov$ $\parallel$ x-axis and $(b_{1y}+r_{1y})r_{2z}=k(b_{1z}+r_{1z})r_{2y}$, where $k$ is an arbitrary constant, then the vector differences $\donev^{C}, \dtwov^C$ are in one of the singular cases. Similar conclusions are readily obtained for rotations around the other two axes by $\pi$ radians.

\textbf{Case D: $\donev \times \dtwov = 0$ \& $\donev \neq 0, \dtwov \neq 0$}
Assume a rotation of $\pi$ radians about the x-axis:
\begin{align*}
    &\ronev=[r_{1x},r_{1y},r_{1z}]^T\\
    &\rtwov=[r_{2x},r_{2y},r_{2z}]^T\\
    &\bonev=[b_{1x},b_{1y},b_{1z}]^T\\
    &\btwov=[b_{2x},b_{2y},b_{2z}]^T\\
    &\ronev^{\Cfr}=[r_{1x},-r_{1y},-r_{1z}]^T\\
    &\rtwov^{\Cfr}=[r_{2x},-r_{2y},-r_{2z}]^T\\
    &\donev^{\Cfr}=\half [ b_{1x}-r_{1x},b_{1y}+r_{1y},b_{1z}+r_{1z}]^T\\
    &\dtwov^{\Cfr}=\half [ b_{2x}-r_{2x},b_{2y}+r_{2y},b_{2z}+r_{2z}]^T
\end{align*}

If $\donev +  [0,r_{1y},r_{1z} ]^T \parallel \dtwov + [0,r_{2y},r_{2z}]^T$, then the vector differences $\donev^{C}, \dtwov^C$ are in one of the singular cases. Similar conclusions are readily obtained for rotations around the other two axes by $\pi$ radians.

\section{Development of the Bias and covariance matrix}\label{sec4}
In this section, an approximation to the second-order estimation error in measurement noise is first presented. Closed-form expressions for the error bias and covariance matrices are then developed, and the covariance matrix is relaxed to a fourth-order approximation to achieve more accurate results.
\subsection{Definitions and notations}
Let $\Dbiv,\Driv$ denote additive errors in the body frame and reference frame vectors, respectively: 
\begin{align}
\biv &= \biv^{t} + \Dbiv, \\
\riv &= \riv^{t} + \Driv, \ 
\end{align}
for $i=1,2$, where the superscript $^t$ denotes the true value of the underlying variable. Let $\qbar$, $\qbar^t$, $\qsca$, and $\qev$ be defined as follows
\begin{align}
\label{s5eq04}
 \qbar &=
 \begin{bmatrix}
 \donev \times \dtwov \\
 \sonev^T \dtwov \\
 \end{bmatrix} \\
 \label{s5eq04a}
 \qsca &= \frac{\qbar}{ |\qbar^t| } \\
\label{s5eq12a}
 \qev &=
\frac{\qbar}{ |\qbar| } \
\end{align}
where $\qbar$ denotes an unnormalized quaternion estimate, $\qsca$ a scaled quaternion estimate, $\qev$ a normalized quaternion estimate, and $\qbar^t$ the true value of $\qbar$. The corresponding estimation errors are defined as follows:
\begin{align}
\label{s5eq06}
 \Dqbar &= \qbar^t - \qbar \\
\label{s5eq06a}
 \Dqsca &= \frac{\Dqbar}{ |\qbar^t| }  \\  
\label{s5eq13}
  \Dqev &= \qv - \qev \\
\label{s5eq14}
 \delqev &= \qv^{-1} \ast \qev \
\end{align} 
where $\qv$, $\ast$, and $\qv^{-1}$ denote the true quaternion, the quaternion composition, and the quaternion inverse. The definition of $\ast$ is provided next for the sake of completeness: given two quaternions $\qv_i$ with vector and scalar parts $\ev_i$ and $q_i$, $i=1,2$, respectively, then 

\begin{align}
\label{s5eq12b}
\qv_1 \ast \qv_2 = 
\begin{bmatrix}
 q_1 \ev_2 + q_2 \ev_1 + \ev_1 \times \ev_2 \\
 \ev_1^T \ev_2 - q_1 q_2 
\end{bmatrix} 
\end{align}

\subsection{Deterministic analysis}
\subsubsection{Exact formulas}
An exact formula for the error $\Dqbar$ is provided as a function of $\Dbv$ and $\Drv$. Using~\eqref{s5eq06} yields the following expression for $\Dqbar$:
\begin{equation}
\begin{aligned}
\label{s5eq15}
\Dqbar = 
\begin{bmatrix}
 [\dtwov^t \times ] \Ddonev + [-\donev^t \times] \Ddtwov - \Ddonev \times \Ddtwov \\
 {- \dtwov^t }^T \Dsonev - {\sonev^t} ^T \Ddtwov - \Dsonev^T \Ddtwov 
\end{bmatrix} \
\end{aligned}
\end{equation}
where
\begin{align}
    & \Dsiv = \half \left( \Dbiv+\Driv \right),\\
    & \Ddiv = \half \left( \Dbiv-\Driv \right),\
\end{align}
for $i=1,2$. Next, we obtain exact expressions for the various errors as a function of $\Dqbar$. Let $\nu$ denote the ratio between the norms of $\qbar$ and $\qbar^t$, i.e.
\begin{align}
\label{s5eq19}
& \nu = \frac{|\qbar^t|}{|\qbar|}, \
\end{align}

Then the following identities are satisfied:
\begin{align}
\label{s5eq21}
& \nu = \left( 1 - 2 \qv^T \Dqsca + |\Dqsca|^2 \right)^{-1/2} \\
\label{s5eq22}
& \Dqev = \nu \Dqsca  + \left ( 1 - \nu \right) \qv \\
\label{s5eq23}
& \delqev = \Uqv + M \Dqev \
\end{align}
where $\Uqv$ and $M$ are defined as follows:
\begin{align}
\label{s5eq24}
& \Uqv = 
\begin{bmatrix}
 \bf{0} \\ 1 \\    
\end{bmatrix} \\
\label{s5eq25}
& M = 
\begin{bmatrix}
 [\ev \times] - q \!\Ithree & \ev  \\
 -\ev^T & - q \\
\end{bmatrix} \
\end{align}
and $\ev$ and $q$ denote the vector and scalar parts of $\qv$, respectively. Equations~\eqref{s5eq15}-\eqref{s5eq25} are a set of exact formulas relating all errors to the underlying measurement errors, and are a useful preliminary to the development of approximations. 

\textbf{The development of Eqs.~\eqref{s5eq21}-~\eqref{s5eq23} are as follows:}

Eq.~\eqref{s5eq21} is developed by considering the squared norm of $\qbar$ first.
\begin{align*}
|\qbar|^2 
& = |\qbar^t - \Dqbar|^2 \\
& = |\qbar^t|^2 - 2 \qbar^t \cdot \Dqbar + |\Dqbar|^2 \
\end{align*}

Dividing by $|\qbar^t|^2$ yields
\begin{align*}
\frac{|\qbar|^2}{|\qbar^t|^2} 
& = 1 - 2 \frac{\qbar^t}{|\qbar^t|} \cdot \frac{\Dqbar}{|\qbar^t|} 
+ \left( \frac{|\Dqbar|}{|\qbar^t|} \right)^2 \\
& = 1 - 2 \qv \cdot \Dqsca + |\Dqsca|^2 \
\end{align*}

Eq.~\eqref{s5eq22} is follows using the definitions of $\qv$, $\Dqsca$, and $\nu$, the developed as follows:
\begin{align*}
\Dqev 
& = \qv - \qev \\
& = \frac{\qbar^t}{|\qbar^t|} - \frac{\qbar}{|\qbar|} \\
& = \frac{\qbar^t}{|\qbar^t|} - \frac{\qbar^t - \Dqbar}{|\qbar|} \\
& = \frac{\Dqbar}{|\qbar|} + 
\left( \frac{1}{|\qbar^t|} - \frac{1}{|\qbar|} \right) \qbar^t \\
& = \frac{|\qbar^t|}{|\qbar|} \frac{\Dqbar}{|\qbar^t|} + 
\left( 1 - \frac{|\qbar^t|}{|\qbar|} \right) \frac{\qbar^t}{|\qbar^t|} \
\end{align*}

Eq.~\eqref{s5eq23} can be get by using the definition of $\delqev= (\qv)^{-1} * \qev $  yields
\begin{align*}
\delqev 
 & = (\qv)^{-1} * \qev \\
 & = (\qv)^{-1} * (\qv - \Dqev) \\
 & = (\qv)^{-1} * \qv - (\qv)^{-1} * \Dqev \\
 & = \Uqv - (\qv)^{-1} * \Dqev \\
 & = \Uqv - \qv^{-1} * \Dqev \
\end{align*}

\subsubsection{Approximation formulas}
We aim here at presenting approximate error expressions that are accurate to second-order in $\Dqbar$, or equivalently in $\Dqsca$. Notice that the expression for the multiplicative error $\delqbar$ is linear in $\Dqsca$. Hence we first study $\nu$ then $\Dqev$. The results are summarized in the following identities:
\begin{align}
\label{s5eq30}
& \nu = 1 + \qv^T \Dqsca - \half \Dqsca^T Q \Dqsca  \\
\label{s5eq31}
& \Dqev = \left[ \Ifour - \qv \qv^T \right] \Dqsca + \left[ \Dqsca  \Dqsca ^T + \half \Dqsca ^T Q \Dqsca  \Ifour \right] \qv \ 
\end{align}
where
\begin{align}
\label{s5eq46}
Q = \Ifour - 3 \qv \qv^T \
\end{align}
Obtaining an approximate expression for $\delqev$ is straightforward since $\delqev$ is linear in $\Dqev$, see~\eqref{s5eq23}. Notice the first-order term in~\eqref{s5eq31}: it is the projection of $\Dqsca$ onto the orthogonal complement to the true quaternion and thus lies in the plane tangent to the unit sphere. This is a well-known effect of quaternion normalization. The second-order term however breaks this property by adding two factors: the first one is along the vector $\Dqsca$ and the second along the true quaternion. To conclude, the current results provide handy formulas for studying the estimation errors as a function of $\Dqbar$. \emph{Notice that these formulas are general, i.e. their scope is not bound to the present estimator.} Nevertheless, using~\eqref{s5eq15} to express $\Dqbar$ yields the sought formulas of $\Dqev$ and $\delqev$ for the proposed estimator.

\textbf{The development of Eqs.~\eqref{s5eq30}-~\eqref{s5eq31} are as follows:}

From Eq.~\eqref{s5eq21}, the ratio $\nu$ is expressed as follows:
\begin{align*}
& \nu = \left( 1 + \epsilon \right)^{-\half} \    
\end{align*}
where 
\begin{align*}
& \epsilon = - 2 \qv^T \Dqsca + |\Dqsca|^2   \ 
\end{align*}

A power series expansion to the second order in $\epsilon$ yields Eq.~\eqref{s5eq21}:
\begin{align*}
\nu 
& = 1 - \half \epsilon + \frac{3}{8} \epsilon^2 \\
& = 1 + \qv^T \Dqsca - \half \Dqsca^T \Dqsca + \frac{3}{2} \Dqsca^T \qv\qv^T \Dqsca \
\end{align*}

Inserting Eq.~\eqref{s5eq30} into Eq.~\eqref{s5eq22}, keeping the second order terms in $\Dqsca$ and rearranging yields Eq.~\eqref{s5eq31}.

\subsection{Random analysis}
The deterministic error analysis lends itself to a random analysis under fairly general assumptions on the underlying vector measurement errors. In particular, we will provide simple expressions for the biases and the covariance matrices that are accurate to the second order and to the fourth order in $\Dbv$ and $\Drv$, respectively.  
\subsubsection{Biases}
Assume that the errors $\Dbiv$ and $\Driv$ are unbiased and mutually uncorrelated random vectors. The unbiasedness assumption holds when the vector measurements are not normalized or when their standard deviations are very small~\cite[p. 204]{mrk1}, and is a common assumption given that biases in sensors are typically evaluated and compensated via calibration. Then the unnormalized errors $\Dqbar$ and $\Dqsca$ are unbiased and the following identities hold:
\begin{align}
\label{s5eq43}
& E\{ \nu \} = 1 - \half \trace(Q\PDqsca) \\
\label{s5eq44}
& E\{ \Dqev \} = \left[ \PDqsca + \half \trace(Q\PDqsca) \Ifour \right] \qv \\
\label{s5eq45}
& E\{ \delqev \} = \Uqv + M \left[ \PDqsca + \half \trace(Q\PDqsca) \Ifour \right] \qv \
\end{align}
where $\PDqsca$ denotes the covariance matrix of $\Dqsca$, i.e. $\PDqsca = E\{\Dqsca \Dqsca^T\}$. These results shed light on the impact of quaternion normalization. The normalized estimation errors $\Dqev$ and $\delqev$ are biased, and their biases are functions of the covariance matrix $\PDqsca$, which will be expressed in the next section. Furthermore,~\eqref{s5eq44} shows that the bias lies close to the true quaternion. This fact is similar to previous findings on unit vector measurements~\cite[5.5.2]{mrk1}, whose bias lies opposite the true vector direction. Yet it is here unclear whether the bias points inward or outward to the unit sphere since the matrix $Q$ is indefinite. Furthermore, the contribution of the term $\PDqsca \qv$ to the bias is not necessarily zero although it is generally small. The proposed estimator is biased, and closed-form expressions for the biases of the additive and multiplicative errors are available. This paves the way for the development of the error covariance matrices.   

\textbf{The development of Eqs.~\eqref{s5eq43}-~\eqref{s5eq45} are as follows:}

Eq.~\eqref{s5eq43} is obtained as follows:
\begin{align*}
E\{\nu\} 
& = E\{ 1 + \qv^T \Dqsca + \half \Dqsca^T \left( 3 \qv \qv^T - \Ifour \right) \Dqsca \} \\
& = 1 + \qv^T E\{  \Dqsca \} + \half E \{ \Dqsca^T \left( 3 \qv \qv^T - \Ifour \right) \Dqsca \} \\
& = 1 + \half \trace  \left[ \left( 3 \qv \qv^T - \Ifour \right) E \{ \Dqsca \Dqsca^T \} \right] \
\end{align*}

Eq.~\eqref{s5eq44} is obtained in a similar manner:
\begin{align*}
E\{ \Dqev \} 
& = E\{ \left[ \Ifour - \qv \qv^T \right] \Dqsca 
+ \left[ \Dqsca  \Dqsca ^T + \half \Dqsca ^T \left( \Ifour - 3 \qv \qv^T \right) \Dqsca  \Ifour \right] \qv \} \\
& = \left[ \Ifour - \qv \qv^T \right] E\{  \Dqsca \} 
+  \left[ E \{ \Dqsca  \Dqsca ^T \} + \half E \{ \Dqsca ^T \left( \Ifour - 3 \qv \qv^T \right) \Dqsca \} \Ifour \right]  \qv  \\
& =  \left\{ \PDqsca + \half 
\trace \left[ \left( \Ifour - 3 \qv \qv^T \right) \PDqsca \right] \Ifour \right\}  \qv  \
\end{align*}

Finally using Eq.~\eqref{s5eq44} in Eq.~\eqref{s5eq23} yields Eq.~\eqref{s5eq45}.

\subsubsection{The covariance matrix $\PDqbar$}

The derivation of $\PDqbar$, the covariance matrix of $\Dqbar$, is straightforward thanks to the simplicity of the estimator. Recalling the expression of $\Dqbar$ from~\eqref{s5eq15} and retaining the linear terms only yields:
\begin{align}
\Dqbar  & = 
\begin{bmatrix}
 0_{3\times3} & E & F \\
 G & 0_{1\times3} & H \\
\end{bmatrix}
\begin{bmatrix}
 \Dsonev \\ \Ddonev \\ \Ddtwov \   
\end{bmatrix} \
\end{align}
where
\begin{align}
& E = [\dtwov^t \times ] \\
& F = [-\donev^t \times ] \\
& G = (-\dtwov^t)^T \\
& H = (-\sonev^t)^T \
\end{align}

Hence the covariance matrix $\PDqbar$ is expressed as follows
\begin{align}
\label{ap5eq10}
& \PDqbar = \begin{bmatrix}
 0_{3\times3} & E & F \\
 G & 0_{1\times3} & H \\
\end{bmatrix}
\begin{bmatrix}
\PDsonev & \PDsonevDdonev & \PDsonevDdtwov  \\
\PDsonevDdonev & \PDdonev & \PDdonevDdtwov \\
\PDsonevDdtwov & \PDdonevDdtwov& \PDdtwov \\
\end{bmatrix}
\begin{bmatrix}
 0_{3\times3} & G^T \\
 E^T & 0_{3\times1} \\
F^T & H^T \end{bmatrix}
\end{align}
where $\PDsonev, \PDdonev, \PDdtwov,\PDsonevDdonev,\PDdonevDdtwov$ denote covariance and cross-covariance matrices that can easily be derived as functions of the measurement errors covariance and cross-covariance matrices $\PDbv, \PDrv, \PDbvDrv$. Using the usual assumptions that the errors from the two VMs are uncorrelated and that $(\Dbonev, \Dronev)$ are uncorrelated yields a simplified expression for $\PDqbar$ as follows:
\begin{align}
\label{s5eq50b}
& \PDqbar =\frac{1}{4}
\begin{bmatrix}
 [\dtwov^t \times ] (\PDbonev+\PDronev) [\dtwov^t \times ]^T + [\donev^t \times] (\PDbtwov+\PDrtwov) [\donev^t \times]^T &  [\donev^t \times ] (\PDbtwov+\PDrtwov) (\sonev^t)  \\
(-\dtwov^t)^T (\PDbonev+\PDronev) [\dtwov^t \times ]^T +  (\sonev^t)^T (\PDbtwov+\PDrtwov) [\donev^t \times ]^T  & (\dtwov^t)^T  (\PDbonev+\PDronev) \dtwov^t  + (\sonev^t)^T  (\PDbtwov+\PDrtwov) \sonev^t  \\
\end{bmatrix} \
\end{align}

The classical case where $\PDbv = \sigma_b^2 (\Ithree - \bv\!\bv^T)$ and $\PDrv = \sigma_r^2 (\Ithree - \rv\!\rv^T)$ can be easily analyzed using~\eqref{s5eq50b}. Assuming for simplicity that the errors $\Dbiv$ and $\Driv$ have the same standard deviation and their covariance matrix is diagonal yields the following expression:
\begin{align}
\label{s5eq50c}
& \PDqbar = 
\frac{\sigma^2}{2}
\begin{bmatrix}
 \sum_{j=1}^2 |\djv^t|^2 \Ithree - \djv^t  (\djv^t)^T & \donev^t \times \sonev^t \\
 (\donev^t \times \sonev^t)^T & |\dtwov^t|^2 + |\sonev^t|^2 \\
\end{bmatrix} \
\end{align}

The particular $\PDqbar$ is developed as follows:
\begin{align*}
& \PDqbar = \frac{\sigma^2}{2}
\begin{bmatrix}
 0_{3\times3} & E & F \\
 G & 0_{1\times3} & H \\
\end{bmatrix}
\begin{bmatrix}
\Ithree & \Ithree & \Zthree  \\
\Ithree & \Ithree & \Zthree \\
\Zthree & \Zthree & \Ithree \\
\end{bmatrix}
\begin{bmatrix}
 0_{3\times3} & G^T \\
 E^T & 0_{3\times1} \\
F^T & H^T \end{bmatrix} \\
& = \frac{\sigma^2}{2}
\begin{bmatrix}
 E E^T + F F^T & F H^T \\
 HF^T & G G^T + H H^T \\
\end{bmatrix} \\
& = \frac{\sigma^2}{2}
\begin{bmatrix}
 \sum_{j=1}^2 |\djv^t|^2 \Ithree - \djv^t  (\djv^t)^T & \donev^t \times \sonev^t \\
 (\donev^t \times \sonev^t)^T & |\dtwov^t|^2 + |\sonev^t|^2 \\
\end{bmatrix}
\end{align*}

\subsubsection{The other covariance matrices}

Let $\PDqbar$, $\PDqev$, $\Pdelqev$  denote the covariance matrices of $\Dqbar$, $\Dqev$, and $\delqev$, respectively. We present results on these covariance matrices for a generic $\PDqbar$, which are summarized next:
\begin{align}
\label{s5eq51}
& \PDqsca = \frac{1}{|\qbar^t|^2} \PDqbar \\
\label{s5eq53}
& \PDqev = \left( \Ifour - \qv \qv^T \right)  \PDqsca \left( \Ifour - \qv \qv^T \right)^T + N \qv \qv^T N^T \\ 
\label{s5eq54}
& \Pdelqev = M \PDqev M^T \ 
\end{align}
where 
\begin{align}
\label{s5eq56}
& N = \PDqsca + \half \trace(Q\PDqsca) \Ifour \
\end{align}
and $M$ and $Q$ are defined in ~\eqref{s5eq25} and~\eqref{s5eq46}, respectively. 

\section{Gaussian Noise}\label{sec5}

From Eq.~\eqref{s5eq31} and Eq.~\eqref{s5eq44}, the deviation is expressed as:
\begin{align*}
&\Dqev - E\{\Dqev\} = \left[ \Ifour - \qv \qv^T \right] \Dqsca + ( [ \Dqsca  \Dqsca ^T + \half \Dqsca ^T Q \Dqsca  \Ifour] - [ \PDqsca + \half \trace(Q\PDqsca) \Ifour] ) \qv \
\end{align*}

The covariance matrix of the normalized quaternion additive error is thus expressed as follows:  
\begin{equation*}
\begin{split}
\PDqev&=E\{(\Dqev - E\{\Dqev\}) (\Dqev - E\{\Dqev\})^T\}\\
&=(\Ifour - \qv \qv^T)  \PDqsca (\Ifour - \qv \qv^T)^T -E\{ \PDqsca \qv \qv^T  \Dqsca \Dqsca^T\} - \frac{1}{2}E\{ \PDqsca \qv \qv^T [\Dqsca^T Q \Dqsca I_4]\}\\
&- \frac{1}{2}E\{ \trace(Q \PDqsca) \Ifour \Dqsca^T \qv \qv^T \Dqsca\} - \frac{1}{4}E\{ \trace(Q \PDqsca) \Ifour \Dqsca^T \qv \trace (Q \Dqsca \Dqsca^T ) \Ifour\}\\
&+E\{ \Dqsca \Dqsca^T \qv \qv^T \Dqsca \Dqsca^T\}+\frac{1}{2}E\{\Dqsca \Dqsca^T \qv \qv^T \trace (Q \Dqsca \Dqsca^T ) \Ifour\}\\
&+\frac{1}{2}E\{ \trace (Q \Dqsca \Dqsca^T ) \Ifour \qv \qv^T  \Dqsca \Dqsca^T\}  +\frac{1}{4}E\{ \trace (Q \Dqsca \Dqsca^T ) \Ifour \qv \qv^T  \trace (Q \Dqsca \Dqsca^T ) \Ifour\}
\end{split}
\end{equation*}

Focus on the fourth-order terms of $\PDqev$, which can be expressed as follows:
\begin{equation*}
\begin{split}
{\PDqev}_{4th}&=E\{ \Dqsca \Dqsca^T \qv \qv^T \Dqsca \Dqsca^T\}+\frac{1}{2}E\{\Dqsca \Dqsca^T \qv \qv^T \trace (Q \Dqsca \Dqsca^T ) \Ifour\}\\
&+\frac{1}{2}E\{ \trace (Q \Dqsca {\Dqsca}^T ) \Ifour \qv {\qv}^T  \Dqsca {\Dqsca}^T\}+\frac{1}{4}E\{ \trace (Q \Dqsca {\Dqsca}^T ) \Ifour \qv {\qv}^T  \trace (Q \Dqsca {\Dqsca}^T ) \Ifour\}\\
&= E\{ {({\Dqsca}^T \qv )}^2 \Dqsca  {\Dqsca}^T \} -3 E\{ ({\Dqsca}^T \qv )^3 \Dqsca {\qv }^T\} +E\{ (\Dqsca^T \Dqsca)  \Dqsca \Dqsca^T \qv  {\qv }^T\}\\
& + \frac{1}{4} E\{ (\Dqsca^T \Dqsca)^2 \qv  {\qv }^T\} -\frac{3}{2} E\{ {({\Dqsca}^T \qv )}^2 (\Dqsca^T \Dqsca) \qv  {\qv }^T \} + \frac{9}{4} E\{ {({\Dqsca}^T \qv )}^4 \qv  {\qv }^T \}
\end{split}
\end{equation*}


Let $\qv =[q_1,q_2,q_3,q_4]^T$ denote the true quaternion, let $\Dqsca=[x_1,x_2,x_3,x_4]^T$ denote the scaled quaternion additive error, and let
$\sigma_i$, $\sigma_{ij}$ denote the standard deviation of $x_i$ and the correlation factor of $x_i$ and $x_j$, respectively. Let's assume that $\{ x_1,x_2,x_3,x_4 \}$ are zero-mean jointly Gaussian random variables, then the following fourth moments~\cite{janssen1988expectation} can be analytically expressed as a function of the second-order deviations:
\begin{align*}
&E\{x_i x_j x_k x_m\}=\sigma_{ij}\sigma_{km}+\sigma_{ik}\sigma_{jm}+\sigma_{im}\sigma_{jk}\\
&E\{x_i^2 x_j x_k\}=\sigma_{i}^2\sigma_{jk}+2\sigma_{ij}\sigma_{ik}\\
&E\{x_i^2 x_j^2 \}=\sigma_{i}^2\sigma_{j}^2+2\sigma_{ij}^2\\
&E\{x_i^3 x_j \}=3\sigma_{i}^2\sigma_{ij}\\
&E\{x_i^4 \}=3\sigma_{i}^4
\end{align*}
  
In the following, we derive analytical expressions for each term of the fourth-order $\PDqev$, The results are summarized in Term \#1 - Term \#6.

\textbf{Term \#1: $E\{ ({\Dqsca}^T \qv )^2 \Dqsca  {\Dqsca}^T \}$ }
\begin{equation*}
    E\{ ({\Dqsca}^T \qv )^2 \Dqsca  {\Dqsca}^T \} = \
\end{equation*}

 =E\resizebox{1\textwidth}{!}{$
$}\qv  {\qv }^T
$

\textbf{Term \#4:  $E\{ {({\Dqsca}^T \Dqsca)}^2 \qv  {\qv}^T\}$ }

\begin{equation*}
    \frac{1}{4} E\{ {({\Dqsca}^T \Dqsca)}^2 \qv  {\qv}^T\}
\end{equation*}

\begin{equation*}
  \begin{aligned}
    &=\frac{1}{4}E\{{x_1}^4+x_2^4+2 x_1^2 x_2^2+x_3^4+2 x_1^2 x_3^2+2x_2^2x_3^2+x_4^4+2 x_1^2 x_4^2+2 x_2^2 x_4^2+2 x_3^2 x_4^2\}\qv  {\qv }^T\\
    &=\frac{1}{4}[3(\sigma_{1}^2+\sigma_{2}^2+\sigma_{3}^2+\sigma_{4}^2)+2(\sigma_{1}^2\sigma_{2}^2+2\sigma_{12}^2+\sigma_{1}^2\sigma_{3}^2+2\sigma_{13}^2+\sigma_{1}^2\sigma_{4}^2+2\sigma_{14}^2+\sigma_{3}^2\sigma_{2}^2\\
    &+2\sigma_{32}^2+\sigma_{4}^2\sigma_{2}^2+2\sigma_{42}^2+\sigma_{3}^2\sigma_{4}^2+2\sigma_{34}^2)] \qv  {\qv }^T
    \end{aligned}  
\end{equation*}

\textbf{Term \#5: $ E\{ ({\Dqsca}^T \qv )^2 (\Dqsca^T \Dqsca) \qv  {\qv }^T  \} $}

\begin{equation*}
    -\frac{3}{2} E\{ ({\Dqsca}^T \qv )^2 (\Dqsca^T \Dqsca) \qv  {\qv }^T \}
\end{equation*}

$=-\frac{3}{2} E
\begin{pmatrix}
\begin{aligned}
&q_1^2x_1^4+q_2^2x_2^4+2q_1q_2x_1x_2^3+q_1^2x_1^2x_2^2+q_2^2x_1^2x_2^2\\
&+2q_1q_2x_1^3x_2+q_3^2x_3^4+2q_1q_3x_1x_3^3+2q_2q_3x_2x_3^3+q_1^2x_1^2x_3^2\\
&+q_3^2x_1^2x_3^2+q_2^2x_2^2x_3^2+q_3^2x_2^2x_3^2+2q_1q_2x_1x_2x_3^2+2q_1q_3x_1^3x_3\\
&+2q_2q_3x_2^3x_3+2q_1q_3x_1x_2^2x_3+2q_2q_3x_1^2x_2x_3+q_4^2x_4^4+2q_1q_4x_1x_4^3\\
&+2q_2q_4x_2x_4^3+2q_3q_4x_3x_4^3+q_1^2x_1^2x_4^2+q_4^2x_1^2x_4^2+q_2^2x_2^2x_4^2\\
&+q_4^2x_2^2x_4^2+2q_1q_2x_1x_2x_4^2+q_3^2x_3^2x_4^2+q_4^2x_3^2x_4^2\\
&+2q_1q_3x_1x_3x_4^2+2q_2q_3x_2x_3x_4^2+2q_1q_4x_1^3x_4+2q_2q_4x_2^3x_4\\
&+2q_1q_4x_1x_2^2x_4+2q_2q_4x_1^2x_2x_4+2q_3q_4x_3^3x_4+2q_1q_4x_1x_3^2x_4\\
&+2q_2q_4x_2x_3^2x_4+2q_3q_4x_1^2x_3x_4+2q_3q_4x_2^2x_3x_4
\end{aligned}
\end{pmatrix}
\qv  {\qv }^T$

$=-\frac{3}{2} 
\begin{pmatrix}
\begin{aligned}
&3q_1^2\sigma_{1}^4+q_1^2(\sigma_{1}^2\sigma_{2}^2+2\sigma_{12}^2+\sigma_{1}^2\sigma_{3}^2+2\sigma_{13}^2+\sigma_{1}^2\sigma_{4}^2+2\sigma_{14}^2)\\
&3q_2^2\sigma_{2}^4+q_2^2(\sigma_{1}^2\sigma_{2}^2+2\sigma_{12}^2+\sigma_{2}^2\sigma_{3}^2+2\sigma_{23}^2+\sigma_{2}^2\sigma_{4}^2+2\sigma_{24}^2)\\
&6q_1q_2(\sigma_{2}^2\sigma_{12}+\sigma_{1}^2\sigma_{12})+2q_1q_2(\sigma_{3}^2\sigma_{12}+2\sigma_{13}\sigma_{23}+\sigma_{4}^2\sigma_{12}+2\sigma_{14}\sigma_{24})\\
&3q_3^2\sigma_{3}^4+q_3^2(\sigma_{1}^2\sigma_{3}^2+2\sigma_{13}^2+\sigma_{2}^2\sigma_{3}^2+2\sigma_{23}^2+\sigma_{3}^2\sigma_{4}^2+2\sigma_{34}^2)\\
&6q_1q_3(\sigma_{3}^2\sigma_{13}+\sigma_{1}^2\sigma_{13})+2q_1q_3(\sigma_{2}^2\sigma_{13}+2\sigma_{12}\sigma_{23}+\sigma_{4}^2\sigma_{13}+2\sigma_{14}\sigma_{34})\\
&6q_2q_3(\sigma_{3}^2\sigma_{23}+\sigma_{2}^2\sigma_{23})+2q_2q_3(\sigma_{1}^2\sigma_{23}+2\sigma_{12}\sigma_{13}+\sigma_{4}^2\sigma_{23}+2\sigma_{24}\sigma_{34})\\
&3q_4^2\sigma_{4}^4+q_4^2(\sigma_{1}^2\sigma_{4}^2+2\sigma_{14}^2+\sigma_{2}^2\sigma_{4}^2+2\sigma_{24}^2+\sigma_{3}^2\sigma_{4}^2+2\sigma_{34}^2)\\
&6q_1q_4(\sigma_{4}^2\sigma_{14}+\sigma_{1}^2\sigma_{14}+2q_1q_4(\sigma_{2}^2\sigma_{14}+2\sigma_{12}\sigma_{24}+\sigma_{3}^2\sigma_{14}+2\sigma_{13}\sigma_{34})\\
&6q_2q_4(\sigma_{4}^2\sigma_{24}+\sigma_{2}^2\sigma_{24})+2q_2q_4(\sigma_{1}^2\sigma_{24}+2\sigma_{12}\sigma_{14}+\sigma_{3}^2\sigma_{24}+2\sigma_{23}\sigma_{34})\\
&6q_3q_4(\sigma_{4}^2\sigma_{34}+\sigma_{3}^2\sigma_{34})+2q_3q_4(\sigma_{1}^2\sigma_{34}+2\sigma_{13}\sigma_{14}+\sigma_{2}^2\sigma_{34}+2\sigma_{23}\sigma_{24})
\end{aligned}
\end{pmatrix}
\qv  {\qv }^T$

\textbf{Term \#6: $ E\{ ({\Dqsca}^T \qv )^4 \qv  {\qv }^T  \} $}

\begin{equation*}
    \frac{9}{4} E\{ ({\Dqsca}^T \qv )^4 \qv  {\qv }^T \}
\end{equation*}

$=\frac{9}{4} E
\begin{pmatrix}
\begin{aligned}
&q_1^4x_1^4+q_2^4x_2^4+4q_1q_2^3x_1x_2^3+6q_1^2q_2^2x_1^2x_2^2+4q_1^3q_2x_1^3x_2+q_3^4x_3^4\\
&+4q_1q_3^3x_1x_3^3+4q_2q_3^3x_2x_3^3+6q_1^2q_3^2x_1^2x_3^2+6q_2^2q_3^2x_2^2x_3^2\\
&+12q_1q_2q_3^2x_1x_2x_3^2+4q_1^3q_3x_1^3x_3+4q_2^3q_3x_2^3x_3+12q_1q_2^2q_3x_1x_2^2x_3\\
&+12q_1^2q_2q_3x_1^2x_2x_3+q_4^4x_4^4+4q_1q_4^3x_1x_4^3+4q_2q_4^3x_2x_4^3\\
&+4q_3q_4^3x_3x_4^3+6q_1^2q_4^2x_1^2x_4^2+6q_2^2q_4^2x_2^2x_4^2+12q_1q_2q_4^2x_1x_2x_4^2\\
&+6q_3+6q_3^2q_4^2x_3^2x_4^2+12q_1q_3q_4^2x_1x_3x_4^2+12q_2q_3q_4^2x_2x_3x_4^2\\
&+4q_1^3q_4x_1^3x_4+4q_2^3q_4x_2x_4+12q_1q_2^2q_4x_1x_2^2x_4+12q_1^2q_2q_4x_1^2x_2x_4\\
&+4q_3^3q_4x_3^3x_4+12q_1q_3^2q_4x_1x_3^2x_4+12q_2q_3^2q_4x_2x_3^2x_4\\
&+12q_1^2q_3q_4x_1^2x_3x_4+12q_2^2q_3q_4x_2^2x_3x_4+24q_1q_2q_3q_4x_1x_2x_3x_4
\end{aligned}
\end{pmatrix}
\qv  {\qv }^T$

$=\frac{9}{4}$
\resizebox{0.95\textwidth}{!}{$\begin{pmatrix}
\begin{aligned}
&3q_1^4\sigma_{1}^4+3q_2^4\sigma_{2}^4+12q_1q_3^3\sigma_{2}^2\sigma_{12}+6q_1^2q_2^2(\sigma_{1}^2\sigma_{2}^2+2\sigma_{12}^2)+12q_1^3q_2\sigma_{1}^2\sigma_{12}+3q_3^4\sigma_{3}^4+12q_1q_3^2\sigma_{3}^2\sigma_{13}\\
&12q_2q_3^3\sigma_{3}^2\sigma_{23}+6q_1^2q_3^2(\sigma_{1}^2\sigma_{3}^2+2\sigma_{13}^2)+6q_2^2q_3^2(\sigma_{2}^2\sigma_{3}^2+2\sigma_{23}^2)+12q_1q_2q_3^2(\sigma_{3}^2\sigma_{12}+2\sigma_{13}\sigma_{23})+12q_1^3q_3\sigma_{1}^2\sigma_{13}\\
&12q_2^3q_3\sigma_{2}^2\sigma_{23}+12q_1q_2^2q_3(\sigma_{2}^2\sigma_{13}+2\sigma_{12}\sigma_{23})+12q_1^2q_2q_3(\sigma_{1}^2\sigma_{23}+2\sigma_{12}\sigma_{13})+3q_4^4\sigma_{4}^4+12q_1q_4^3\sigma_{4}^2\sigma_{14}\\
&12q_2q_4^3\sigma_{4}^2\sigma_{24}+12q_3q_4^3\sigma_{4}^2\sigma_{34}+6q_1^2q_4^2(\sigma_{1}^2\sigma_{4}^2+2\sigma_{14}^2)+6q_2^2q_4^2(\sigma_{2}^2\sigma_{4}^2+2\sigma_{24}^2)+12q_1q_2q_4^2(\sigma_{4}^2\sigma_{12}+2\sigma_{14}\sigma_{24})\\
&6q_3^2q_4^2(\sigma_{3}^2\sigma_{4}^2+2\sigma_{34}^2)+12q_1q_3q_4^2(\sigma_{4}^2\sigma_{13}+2\sigma_{14}\sigma_{34})+12q_2q_3q_4^2(\sigma_{4}^2\sigma_{23}+2\sigma_{24}\sigma_{34})+12q_1^3q_4\sigma_{1}^2\sigma_{14}\\
&12q_2^3q_4\sigma_{2}^2\sigma_{24}+12q_1q_2^2q_4(\sigma_{2}^2\sigma_{14}+2\sigma_{12}\sigma_{24})+12q_1^2q_2q_4(\sigma_{1}^2\sigma_{24}+2\sigma_{12}\sigma_{14})+12q_3^3q_4\sigma_{3}^2\sigma_{34}+12q_1q_3^2q_4(\sigma_{3}^2\sigma_{14}+2\sigma_{13}\sigma_{34})\\
&12q_2q_3^2q_4(\sigma_{3}^2\sigma_{24}+2\sigma_{23}\sigma_{34})+12q_1^2q_3q_4(\sigma_{1}^2\sigma_{34}+2\sigma_{13}\sigma_{14})+12q_2^2q_3q_4(\sigma_{2}^2\sigma_{34}+2\sigma_{23}\sigma_{24})+24(\sigma_{12}\sigma_{34}+\sigma_{13}\sigma_{24}+\sigma_{14}\sigma_{23})\\
\end{aligned}
\end{pmatrix}$}$\qv  {\qv }^T$

\section{Conclusion}\label{sec7}
This paper introduced a single-frame quaternion estimator processing two vector observations. The exceedingly simple structure of the estimator was instrumental in obtaining analytical expressions and clear geometrical insights. The singular cases are analyzed and the corresponding valid rotation solutions are given, another sequential rotation method is also introduced to address the singularity. An in-depth study yielded analytical expressions for the biases and covariance matrices of the estimation error. The formulas are second-order approximations to the measurement noises. In particular, fourth-order approximation formulas were developed for the Gaussian case. It not only provides increased accuracy but also alleviates issues related to singularity. All developments were performed in the four-dimensional quaternion algebra rather than in three dimensions.


\end{document}